\documentclass[12pt]{article}

\newcommand{\commentstarts}{\begin{centering}
\hspace{-1pt}\vrule\vrule
\begin{minipage}[t]{0.03\linewidth}
\hspace{0.025\linewidth}
\end{minipage}
\begin{minipage}[t]{0.95\linewidth}}
\newcommand{\commentends}{\end{minipage}
\end{centering}
\vspace{7pt}
}

\usepackage{graphicx}
\usepackage{alltt}
\usepackage{amsmath}
\usepackage{amssymb}
\usepackage{hyperref}

\begin{document} 

\newcounter{Theorems}
\setcounter{Theorems}{0}

\newcounter{Definitions}
\setcounter{Definitions}{0}

\begin{titlepage}
\begin{flushright}

\end{flushright}

\begin{center}
{\Large\bf $ $ \\ $ $ \\
Integration over families of Lagrangian submanifolds in BV formalism
}\\
\bigskip\bigskip\bigskip
{\large Andrei Mikhailov${}^{\dag}$}
\\
\bigskip\bigskip
{\it Instituto de F\'{i}sica Te\'orica, Universidade Estadual Paulista\\
R. Dr. Bento Teobaldo Ferraz 271, 
Bloco II -- Barra Funda\\
CEP:01140-070 -- S\~{a}o Paulo, Brasil\\
}

\vskip 1cm
\end{center}

\begin{abstract}
Gauge fixing is interpreted in BV formalism as a choice of Lagrangian submanifold in an odd 
symplectic manifold (the BV phase space). A natural construction defines an integration
procedure on families of 
Lagrangian submanifolds. In string perturbation theory, the moduli space integrals of higher 
genus amplitudes can be interpreted in this way. We discuss the role of gauge symmetries in this 
construction. We derive the conditions which should be imposed on gauge symmetries for the 
consistency of our integration procedure. We explain how these conditions behave under the
deformations of the worldsheet theory. In particular, we show that integrated vertex operator is 
actually an inhomogeneous differential form on the space of Lagrangian submanifolds.
\end{abstract}

\vfill
{\renewcommand{\arraystretch}{0.8}%
\begin{tabular}{rl}
${}^\dag\!\!\!\!$ 
& 
\footnotesize{on leave from Institute for Theoretical and 
Experimental Physics,}
\\    
&
\footnotesize{ul. Bol. Cheremushkinskaya, 25, 
Moscow 117259, Russia}
\\
\end{tabular}
}

\end{titlepage}

\tableofcontents 
\section{Introduction}

Many physical theories have BRST-like structure. This means that there is a nilpotent fermionic 
symmetry $Q$, and the space of physical states is the cohomology of $Q$.
Such theories come with an equivalence relation. The theory with the action $S$ is 
equivalent to the theory with the action $S + Q\Psi$ for any operator $\Psi$. Physically meaningful 
quantities should be invariant, {\it i.e.} should be the same for any two equivalent theories. 
An important question is, how can we obtain such invariants?
One way to obtain an invariant is to take the path integral:
\begin{equation} \int_{\rm\scriptstyle {path\atop integral}} [d\phi] \;{\cal O}_1\cdots{\cal O}_n\; e^{{1\over \hbar}S[\phi]} \end{equation}
where ${\cal O}_j$ are some BRST-closed operators.
But actually there are other possibilities \cite{Schwarz:2000ct,Mikhailov:2016myt}, which we will now describe.

Consider the whole equivalence class of theories. It is infinite-dimensional, because 
$S\simeq S + Q\Psi$ where $\Psi$ could be more or less arbitrary functional. 

Let us choose a basis $\{\Psi_a\}$; the space of BRST trivial deformations is parametrized by the 
coordinates $x^a$:
\begin{equation}\label{DeformedS} S(x) = S + \sum_a x^a Q\Psi_a \end{equation}
Let us call this equivalence class $\cal M$. It turns out that the following pseudo-differential 
form on $\cal M$ is closed:
\begin{align}\label{NaiveOmega}
\Omega \;=\; & \int_{\rm\scriptstyle {path\atop integral}} [d\phi] \exp\left(S[\phi] \;+\; \sum_a \Psi_a[\phi] dx^a \right)
\end{align}
We want to obtain BRST invariants by integrating $\Omega$ over some closed cycles \cite{Schwarz:2000ct}.

\commentstarts{\small
This construction (and related) was also used in \cite{Bonechi:2011um,Cattaneo:2015vsa}, and on manifolds with a boundary in \cite{Cattaneo:2015vsa,Cattaneo:2016zrn}.
}\commentends

\noindent
This procedure was used in \cite{Schwarz:2000ct} to define string amplitudes\footnote{BV formalism was applied to string worldsheet theory in \cite{Craps:2005wk}}. We have to integrate over the 
moduli space of metrics on a genus $g$ Riemann surface. When we vary the metric, the variation
of the worldsheet action is $Q$-exact:
\begin{equation}
T_{\alpha\beta} = Q b_{\alpha\beta}
\end{equation}
In this case $\Psi_a dx^a\;=\;b_{\alpha\beta}dg^{\alpha\beta}$ and Eq. (\ref{NaiveOmega}) gives the standard string measure:
\begin{equation}\label{OmegaString}
\Omega \;=\; \int [d\phi] \exp\left(
    S[\phi] + \int_{\rm\scriptstyle worldsheet} b_{\alpha\beta} dg^{\alpha\beta}
\right)
\end{equation}
where $\phi$ are the fields of the worldsheet sigma-model. In this approach, the choice of a
metric $g^{\alpha\beta}$ should be understood as a choice of a representative in the class of physically
equivalent theories (a choice of gauge fixing).

We would like to integrate over the {\em moduli space} of metrics modulo diffeomorphisms. 
However, the form $\Omega$ given by Eq. (\ref{OmegaString}) is {\em not automatically} base. Although it is true
that it is diffeomorphism-invariant, it is generally speaking not horizontal and therefore not
a base form. Indeed, changing $dg^{\alpha\beta}$ to $dg^{\alpha\beta} + 2\nabla^{(\alpha}\xi^{\beta)}$ results in the same expression for 
$\Omega$ only when:
\begin{equation}\label{NablaBIsZero}
\nabla^{\alpha}b_{\alpha\beta} = 0
\end{equation}
Even in the case of standard bosonic string, this is only true on-shell, leading to restrictions
on allowed operator insertions. For a general string theory sigma-model (such as pures spinor 
formalism in a curved background \cite{Berkovits:2001ue}) we would like to relax the condition (\ref{NablaBIsZero}) by
allowing $\nabla^{\alpha} b_{\alpha\beta}$ to be $Q$ of something. Moreover, for the worldsheet theories of 
the topological type, such as the pure spinor formalism \cite{Berkovits:2001ue}, a deformation of the metric is
not necessarily better than any other $Q$-exact deformation of the action. We want to extend the 
definition of $\Omega$ to the space of {\em all} BRST-trivial deformations. But with all these 
generalizations, we still want $\Omega$ to be a {\em base form} with respect to the worldsheet 
diffeomorphisms.

A construction of base $\Omega$ was suggested in \cite{Mikhailov:2016myt}. The purpose of this paper is to further develop
this construction, and to fill in some technical details.

It appears that the right interpretation of $\Omega$ is a pseudo-differential form on the space of 
Lagrangian submanifolds in BV phase space, which we denote $\rm LAG$. \marginpar{$\rm LAG$}
We will give the definition in Section \ref{sec:FormOmega}. The space $\rm LAG$ does not have topologically 
nontrivial closed integration cycles. One can integrate over 
a non-compact cycle, extending to infinity, if $\Omega$ is rapidly decreasing. But this is 
not the case in string theory. Therefore we have to consider the factorspace of $\rm LAG$ over 
the action of the group of worldsheet diffeomorphisms. In our approach, the action of the 
group of diffeomorphisms on the BV phase space should come as part of the definition of the
string worldsheet theory. In more general applications (beyond string theory) we may use,
in place of diffeomorphisms, some other symmetry group. We develop the formalism for general
symmetry group. For our construction of base $\Omega$ to work, the action of the symmetry group 
on the BV phase space should satisfy certain properties, which we describe in Section \ref{sec:EquivariantOmega}. 
One special case is when BV formalism comes from BRST formalism; we discuss this in Section \ref{sec:BRST}. 
Another special case is topologically twisted $N=2$ superconformal theory (Section \ref{sec:N2}). 

In string theory, one considers the worldsheet theory along with its nontrivial deformations. 
These deformations are called ``integrated vertex operators''\footnote{BRST-trivial deformations do not change the physical worldsheet 
theory.  BRST-nontrivial deformations, on the other hand, do change the worldsheet theory, 
but do not change ``the string theory''. They just change the background.}.  In Section \ref{sec:IntegratedVertexOperators} we study the 
corresponding deformation of the string measure. We show that this deformation will generally 
speaking involve mixing between the components of $\Omega$ of different degrees. A closely related 
concept is unintegrated vertex operators; we will consider them in Section \ref{sec:UnintegratedVertexOperators}.
In Section \ref{sec:WorldsheetWithBoundary} we discuss the case of worldsheet with boundary. This could be useful for the
off-shell formulation of string theory \cite{Sen:2014pia}. 

\section{Brief review of the BV formalism}\label{sec:ReviewBV}
The ``BV phase space'' is a supermanifold $M$ with a non-degenerate odd 2-form $\omega$ satisfying 
$d\omega=0$. It is {\bf not true} that $\omega^{-1}$ defines an antisymmetric bivector. 
(The symmetry properties of $\omega^{-1}$ are not of a bivector, see Appendix \ref{sec:PoissonStructure}.) 
Instead, it defines a second order odd differential operator $\Delta$ 
on the so-called ``half-densities'' \cite{Khudaverdian:1999}. A half-density is a geometrical object which 
transforms under the change of variables as a square root of the volume element. 
A half-density defines a measure on every Lagrangian submanifold $L\subset M$.

\subsection{Canonical transformations}\label{sec:CanonicalTransformations}
Let $M$ denote the BV phase space. We say that a 
vector field is Hamiltonian if it preserves the odd symplectic form. Let us assume that $M$ is 
simply-connected. Then Hamiltonian vector fields are of the form $\{H,\_\}$.
Let $\bf g$ denote the Lie algebra of Hamiltonian vector fields.  \marginpar{$\bf g$}
The commutator is defined as the standard commutator of vector fields:
\begin{equation}
[\{H_1,\_\},\{H_2,\_\}] = \{\{H_1,H_2\},\_\}
\end{equation}
Let $G$ denote the Lie supergroup of canonical transformations of $M$.
\marginpar{$G$}
Its Lie superalgebra is opposite to the Lie superalgebra of Hamiltonian vector fields\footnote{If we consider the space of vector fields as a Lie algebra of the group of diffeomorphisms,
then the usual definition of the Lie algebra bracket has opposite sign to the standard
commutator of vector fields}:
\begin{equation}\label{OppositeLieAlgebra}
{\bf g} = - \mbox{Lie}(G)
\end{equation}
Let $C^{\infty}M$ denote the space of smooth functions on $M$. 
Notice that $\Pi C^{\infty}M$ is a Lie superalgebra under the Poisson bracket. It is a central extension of $\bf g$: \marginpar{$\widehat{\bf g}$}
\begin{equation}\label{DefHatG} \widehat{\bf g} = \Pi C^{\infty}M \end{equation}
Let $\widehat{G}$ denote the corresponding central extension of $G$. 
\marginpar{$\widehat{G}$}
It can be realized as the group of automorphisms of some contact manifold --- see
Appendix \ref{sec:Quantomorphisms}.

\subsection{Canonical operator}\label{sec:CanonicalOperator}

\subsubsection{Definition of $\Delta_{\rm can}$}
Any half-density $\rho_{1\over 2}$ defines a measure on a Lagrangian submanifold $L$ \cite{Schwarz:1992nx,Schwarz:1992gs}, which we will denote 
$\left.\rho_{1\over 2}\right|_L$, or sometimes just $\rho_{1\over 2}$. Given a smooth function $H$, let us consider the variation of 
$\int_L\rho_{1\over 2}$ under the variation of $L$ specified by the Hamiltonian vector field $\xi_H$ corresponding 
to $H$. It can only depend on the restriction of $H$ on $L$. Therefore it should be of the form:
\begin{equation}\label{MeasureMu} \delta_{\{H,\_\}} \int_L\left.\rho_{1\over 2}\right|_L \; = \,\int_L H\; \mu_L[\rho_{1\over 2}] \end{equation}
where $\mu_L[\rho_{1\over 2}]$ is some integral form on $L$ (which of course depends on $\rho_{1\over 2}$). 
There exists some half-density on $M$, which we will denote $\Delta_{\rm can}\rho_{1\over 2}$, such that:\marginpar{$\Delta_{\rm can}$}
\begin{equation} \mu_L[\rho_{1\over 2}] = - \left.\left(\Delta_{\rm can}\rho_{1\over 2}\right)\right|_L \end{equation}
In other words:

\refstepcounter{Theorems}
\paragraph     {Theorem \arabic{Theorems}:\label{theorem:DeltaCanonical}} 
given a half-density $\rho_{1\over 2}$, there exists another half-density $\Delta_{\rm can}\rho_{1\over 2}$, such that for any 
$H\in\mbox{Fun}(M)$ and any Lagrangian $L\subset M$:
\begin{equation}
\label{DeltaCanonical} \delta_{\{H,\_\}} \int_L\left.\rho_{1\over 2}\right|_L \; = \, - \int_L H\; \left.\left(\Delta_{\rm can}\rho_{1\over 2}\right)\right|_L 
\end{equation}
Eq. (\ref{DeltaCanonical}) is the definition of $\Delta_{\rm can}$. We will give a proof in Appendix \ref{sec:ProofOfTheoremDef}.

\subsubsection{Relation between $\Delta_{\rm can}$ and Lie derivative}
Let us fix two functions $F\in \mbox{Fun}(M)$ and $\Psi\in \mbox{Fun}(M)$.
Let us suppose that $\Psi$ is odd. Then:
\begin{equation} \{F,\Psi\} = - \{\Psi,F\} \end{equation}
For any Lagrangian submanifold $L\subset M$, let us consider:
\begin{align}
  \phantom{=\;} \int_L \left(\Psi{\cal L}_{\{F,\_\}}\rho_{1\over 2} + F{\cal L}_{\{\Psi,\_\}}\rho_{1\over 2}\right) \;=\;
  \int_L \left({\cal L}_{\{F,\_\}}(\Psi\rho_{1\over 2}) + {\cal L}_{\{\Psi,\_\}}(F\rho_{1\over 2})\right)\;=\;
\nonumber\\
\;=\; \delta_{\{F,\_\}}\int_L \Psi\rho_{1\over 2} + \delta_{\{\Psi,\_\}}\int_L F\rho_{1\over 2} \;=\; -\int_L F\Delta_{\rm can}(\Psi\rho_{1\over 2}) - \int_L \Psi\Delta_{\rm can}(F\rho_{1\over 2})\label{f-h-minus-h-f}
\end{align}
Consider the case when the restriction of $\Psi$ to $L$ is zero. Then Eq. (\ref{f-h-minus-h-f}) implies that the
restriction of ${\cal L}_{\{\Psi,\_\}}\rho_{1\over 2}$ on such $L$ is equal to $-\Delta_{\rm can} (\Psi\rho_{1\over 2})$. We will use $F$ as a ``test
function'' and assume that $F$ has  compact support, contained in a sufficiently small
open superdomain $U\subset M$.

\commentstarts{\small
  A superdomain $U$ of dimension $m|n$ is defined \cite{BernsteinLecture1} through its algebra of functions:
  $C^{\infty}(U) = C^{\infty}(U_{\rm rd})\otimes \Lambda^{\bullet}{\bf R}^n$ where $U_{\rm rd}\subset {\bf R}^{m}$ is an open set and
  $\Lambda^{\bullet}{\bf R}^n = {\bf R}[\theta^1,\ldots,\theta^n]$ is the Grassmann algebra built on fermionic variables $\theta^1,\ldots,\theta^n$.
  Both $F$ and $\Psi$ are functions of $x^1,\ldots,x^m,\theta^1,\ldots,\theta^n$.
  }\commentends

\noindent
The submanifold $U_0\subset U$ given by the equation $\Psi=0$ contains sufficiently many Lagrangian
submanifolds, in the following sense: if the restriction of a density on any Lagrangian
submanifold contained in $U_0$ is zero, then the density is zero everywhere\footnote{If we were working with ordinary (not super) manifolds, we would say that through
  every pointof $U$ passes at least one Lagrangian submanifold fully contained in $U_0$.} on $U_0$.

\commentstarts{\small
  Indeed, when $U$ is small enough, we can consider the space of trajectories of $\{\Psi,\_\}$ on $U_0$.
  It is an odd symplectic manifold (the odd analogue of the Hamiltonian reduction). It has
  sufficiently many Lagrangian submanifolds, in the above sense. They lift to
  Lagrangian submanifolds in $U$.
  }\commentends

\noindent
Therefore Eq. (\ref{f-h-minus-h-f}) implies that on $U_0$: ${\cal L}_{\{\Psi,\_\}}\rho_{1\over 2} = -\Delta_{\rm can} (\Psi\rho_{1\over 2})$. To extend this formula
from $U_0$ to the whole $U$, let us consider the superdomain $\hat{U} = {\bf R}^{0|1}\times U$; the fermionic
coordinate of ${\bf R}^{0|1}$ will be denoted $\zeta$. Consider the subspace of $\hat{U}_0\subset \hat{U}$ given by the equation
$\zeta - \Psi(x,\theta) = 0$. It has sufficiently many maximally isotropic submanifolds. Then the same
computation as in Eq. (\ref{f-h-minus-h-f}) gives:
\[
   {\cal L}_{\{\Psi,\_\}} \rho_{1\over 2} = - \Delta_{\rm can}( (\Psi-\zeta)\rho_{1\over 2} ) + (\Psi - \zeta)X =
   - \Delta_{\rm can}(\Psi\rho_{1\over 2}) - \zeta \Delta_{\rm can} \rho_{1\over 2} + (\Psi - \zeta)X
\]
where $X$ is some funcion on $\hat{U}$. But ${\cal L}_{\{\Psi,\_\}} \rho_{1\over 2}$ by definition does not depend on $\zeta$. Therefore
$X=-\Delta_{\rm can}\rho_{1\over 2}$. This implies, for odd $\Psi$:
\begin{equation}
{\cal L}_{\{\Psi,\_\}} \rho_{1\over 2} = - \Delta_{\rm can}( \Psi\rho_{1\over 2} ) - \Psi\Delta_{\rm can}\rho_{1\over 2}
\end{equation}
If instead of odd $\Psi$ we consider some even $H$, then this argument does not work,
because when $\{H,H\}\neq 0$ there are no Lagrangian submanifolds
contained in level sets of $H$. But, given some odd $\Psi$ and a constant Grassmann parameter $\varepsilon$, we can apply
the argument to the odd Hamiltonian $\Psi + \varepsilon H$. Considering the coefficient of $\varepsilon$ proves that for
even $H$:
\begin{equation}
{\cal L}_{\{H,\,\_\}}\rho_{1\over 2} = \Delta_{\rm can}\left( H\rho_{1\over 2} \right) - H\Delta_{\rm can}\rho_{1\over 2} 
\end{equation}
The formula which works for both even and odd $H$ is:
\begin{equation}\label{ViaLieDerivative}
{\cal L}_{\{H,\,\_\}}\rho_{1\over 2} = (-)^{\bar{H}}\Delta_{\rm can}\left( H\rho_{1\over 2} \right) - H\Delta_{\rm can}\rho_{1\over 2} 
\end{equation}

\subsubsection{The canonical operator is nilpotent}
Indeed, since the definition of $\Delta_{\rm can}$ is geometrically natural, it automatically commutes with canonical transformations and therefore for any $H\in \mbox{Fun}(M)$:
\begin{equation} [\Delta_{\rm can}, {\cal L}_{\{H,\_\}}]\rho_{1\over 2} \;=\;0 \end{equation}
Comparing this with Eq. (\ref{ViaLieDerivative}) we conclude that $\Delta_{\rm can}^2$ commutes with multiplication by an
arbitrary function. Therefore $\Delta_{\rm can}^2$ is multiplication by a function. Morever, since
$\Delta_{\rm can}^2$ commutes with any canonical transformation, this function should be a constant. Taking into
account that for any half-density $\rho_{1\over 2}$ and any Lagrangian submanifold $L$: $\int_L \Delta_{\rm can}^2 \rho_{1\over 2} = 0$, we conclude
that the this constant is zero, {\it i.e.}:
\begin{equation} \Delta_{\rm can}^2 = 0 \end{equation}

\subsubsection{Odd Laplace operator on functions}	
Given a half-density $\rho_{1/2}$ we can define the odd Laplace operator on functions
as follows:\marginpar{$\Delta_{\rho_{1/2}}$}
\begin{align}
(\Delta_{\rho_{1/2}} F)\rho_{1/2} \;=\; &
\Delta_{\rm can}(F\rho_{1/2}) - (-)^{\bar{F}}F\Delta_{\rm can}\rho_{1\over 2} \;=\;
(-)^{\bar{F}}{\cal L}_{\{F,\_\}}\rho_{1\over 2}
\end{align}
We will often abbreviate:\marginpar{$\Delta$}
\begin{equation}
\Delta F \;=\; \Delta_{\rho_{1/2}} F 
\end{equation}
when there is some obvious implicit choice of half-density.

We will now derive a formula for $\Delta$ of the product of two functions. 
But first we have to discuss some properties of Lie derivative which are independent of its relation to $\Delta_{\rm can}$.

\paragraph     {Some properties of the Lie derivative of a half-density}

Consider a vector field $v$ on $M$, and the corresponding 1-parameter group of diffeomorphisms $g^t$.
Let us think of a half-density $\rho_{1\over 2}$ as a function of $x$ and $\bf E$, where $x$ is a point of $M$ and $\bf E$
a basis in $T_xM$, depending on $\bf E$ in the following way:
\begin{equation}
\rho_{1\over 2}(x,A{\bf E}) = \left(\mbox{SDet}A\right)^{1/2}\rho_{1\over 2}(x,{\bf E})
\end{equation}
By definition, the Lie derivative of $\rho_{1\over 2}$ along $v\in \mbox{Vect}(M)$ is:
\begin{equation}
\left({\cal L}_v \rho_{1\over 2}\right)(x,{\bf E}) = \left.{d\over dt}\right|_{t=0} \rho(g^tx, g^t_*{\bf E})
\end{equation}
Let us multiply $v$ by a function $f\in\mbox{Fun}(M)$ such that $f(x)=0$. The flux of $fv$ 
preserves the point $x$, and we have:
\begin{align}
& \left({\cal L}_{fv} \rho_{1\over 2}\right)(x,{\bf E}) =
(-)^{\bar{f}\bar{v}}\left.{d\over dt}\right|_{t=0} \rho\left(x, \exp(t v\otimes df){\bf E}\right) =
{(-)^{\bar{f}\bar{v}} \over 2}({\cal L}_vf)\rho_{1\over 2}(x,{\bf E})
\nonumber
\end{align}
This implies that for any $f\in \mbox{Fun}(M)$ and $v\in \mbox{Vect}(M)$:
\begin{equation}
{\cal L}_{fv} \rho_{1\over 2} = f{\cal L}_v \rho_{1\over 2} + (-)^{\bar{f}\bar{v}} {1\over 2} ({\cal L}_vf)\rho_{1\over 2}
\end{equation}
In particular:
\begin{align}
{\cal L}_{\{FH,\_\}}\rho_{1\over 2}\;=\; &
{\cal L}_{F\{H,\_\}}\rho_{1\over 2} + (-)^{\bar{F}\bar{H}}{\cal L}_{H\{F,\_\}}\rho_{1\over 2} \;=\;
\nonumber\\   
\;=\; &
F{\cal L}_{\{H,\_\}}\rho_{1\over 2} + (-)^{\bar{F}\bar{H}}H{\cal L}_{\{F,\_\}}\rho_{1\over 2} + (-)^{(\bar{H}+1)\bar{F}} \{H,F\}\rho_{1\over 2}\label{Lie-derivative-along-FH}
\end{align}
\paragraph     {Odd Laplace operator of product of functions}
Eqs. (\ref{ViaLieDerivative}) and (\ref{Lie-derivative-along-FH}) imply:
\begin{equation}\label{BVStructure}
\Delta_{\rho_{1/2}}(XY\,) = (\Delta_{\rho_{1/2}} X)Y + (-)^{\overline{X}}X\Delta_{\rho_{1/2}} Y + (-)^{\overline{X}}\{X,Y\,\}
\end{equation}
Notice that the odd Poisson bracket measures the deviation of $\Delta_{\rho_{1/2}}$ from being a 
differentiation of $\mbox{Fun}(M)$.

\subsection{Master Equation}
We will always assume that $\rho_{1\over 2}$ satisfies the Master Equation:
\begin{equation} \Delta_{\rm can}\rho_{1\over 2} = 0 \end{equation}
Under this assumption, the operator $\Delta_{\rho_{1/2}}$ is nilpotent:
\begin{equation} \Delta_{\rho_{1/2}}^2 = 0 \end{equation}

\subsection{Moment map}\label{sec:MomentMap}
Let $G$ denote the group of canonical transformations of $M$. Let us denote $G_L$ and $G_R$ the group 
of left and right shifts on $G$ (both $G_L$ and $G_R$ are naturally isomorphic to $G$). Both left and 
right shifts naturally lift to $\Pi TG$. Let $\bf g$ be the Lie algebra corresponding to $G$.

We consider the right-invariant differential form on $\widehat{G}$ with values in $\widehat{\bf g}$, which is denoted 
$d\hat{g}\hat{g}^{-1}$. Because of (\ref{DefHatG}), we can consider it as a differential 1-form with values in
$\Pi C^{\infty}M$. There is some invariance under the left and right shifts, which can be summarized 
as follows: \marginpar{$d\widehat{g}\widehat{g}^{-1}$}
\begin{align}
d\widehat{g}\widehat{g}^{-1}\;\in\; & C^{\infty}\left( (\Pi T\widehat{G})\times_{\widehat{G}_L\times \widehat{G}_R} M \right)\label{moment-map-type}
\\
d\widehat{g}\widehat{g}^{-1} \;=\; & (d\widehat{g}\widehat{g}^{-1})^A{\cal H}_A\label{moment-map-in-coordinates}
\\
d_{(g)} F(gx) \;=\; & \{d\widehat{g}\widehat{g}^{-1}\,,\,F\,\}(gx)\label{moment-map-derivative-of-f}
\end{align}
where the action of $\widehat{G}_L\times \widehat{G}_R$ on $\Pi T\widehat{G}$ is induced from the following action on $\widehat{G}$:
\begin{equation} (g_L,g_R) \; g \;= g_Lgg^{-1}_R \end{equation}
The action of $\widehat{G}_L\times \widehat{G}_R$ on $M$ is:
\begin{equation}
(g_L,g_R) \; m \;= g_Lm
\end{equation}
The $\widehat{G}_R$-invariance of $d\widehat{g}\widehat{g}^{-1}$ follows immediately from the definitions. The $\widehat{G}_L$-invariance, at the
infinitesimal level, follows from the following formula, where it is enough to assume that $\{\xi^B{\cal H}_B\,,\,\_\}$ 
is an even vector field:
\begin{align}
\left.{d\over dt}\right|_{t=0}(d\widehat{g}\widehat{g}^{-1})^A{\cal H}_A\left(e^{-t\{\xi^B{\cal H}_B\,,\,\_\}}x\right)
\;=\;&
-(d\widehat{g}\widehat{g}^{-1})^A\xi^B\left\{{\cal H}_B, {\cal H}_A\right\}(x)\;=\;
\nonumber\\     
\;=\;&-\left[\xi,d\widehat{g}\widehat{g}^{-1}\right]^A{\cal H}_A(x)
\end{align}
The moment map satisfies the Maurer-Cartan equation: 
\begin{equation}\label{MaurerCartan}
d(d\widehat{g}\widehat{g}^{-1}) = 
- {1\over 2}\{d\widehat{g}\widehat{g}^{-1}\,,\,d\widehat{g}\widehat{g}^{-1}\}
\end{equation}
\commentstarts{\small
Minus signs are related to the minus sign in Eq. (\ref{OppositeLieAlgebra}). The group of diffeomorphism is
{\em opposite} to the Lie group whose Lie algebra is vector fields. Therefore the minus sign 
in $e^{-t\{\xi^B{\cal H}_B,\_\}}x$ and the minus sign in Eq. (\ref{MaurerCartan}).
}
\commentends

\section{Canonical operator is ill-defined in field-theoretic context}
The considerations of Section \ref{sec:ReviewBV} are valid when the BV phase space $M$ is a finite-dimensional 
supermanifold. But in the field theory context, $\Delta_{\rm can}$ is usually ill-defined. We use the BV formalism 
for the worldsheet sigma-model, {\it i.e.} in the field-theoretic context.

We hope that it is possible to regularize $\Delta_{\rm can}$, maybe as in \cite{CostelloRenormalization}. In many cases the string 
worldsheet theory is essentially free (quadratic Lagrangian) and the path integral can be 
computed exactly, allowing the explicit verification. In other cases, there is a natural 
small parameter and the computations can  be verified order by order in perturbation theory.

In string theory models, we need to act by $\Delta$ on the $b$-ghost.
Therefore we need ad additional assumption; roughly speaking,
the symmetry generated by the $b$-ghost should be nonanomalous.
The validity of this assumption is model-dependent; it should be imposed in addition to
the absense of BRST anomaly.

In this paper, we just use $\Delta_{\rm can}$ in formulas as if it were well-defined.

\section{Form $\Omega$}\label{sec:FormOmega}

\subsection{Summary}\label{sec:OmegaSummary}
We want to define some pseudo-differential form $\Omega$ which can serve as {\em string measure}.
There are several closely related definitions:
\begin{enumerate}
\item 
{\bf PDF on the group of canonical transformations.} In Section \ref{sec:OmegaOnPiTGxLAG} we will start by constructing $\Omega$ as a PDF on $\widehat{G}$. The definition actually 
depends on the choice of a fixed Lagrangian submanifold $L\subset M$. 
Strictly speaking, we can characterize $\Omega$ as a map of the following type:
\begin{equation}\label{TypeOfOmegaOnGHat} \Omega\;:\;{\rm LAG} \rightarrow \mbox{Fun}(\Pi T\widehat{G}) \end{equation}
(which associates a PDF on $\widehat{G}$ to every Lagrangian submanifold).
\item {\bf PDF on the space of Lagrangian submanifolds.} There is a natural map:
\begin{equation} \widehat{G}\times {\rm LAG}\;\rightarrow\;{\rm LAG} \end{equation}
coming from the action of $G$ on $M$. A natural question is, does $\Omega$ descend to a PDF 
on $\rm LAG$? The answer is essentially ``yes'', although some minor modifications are needed. 
In Section \ref{sec:DescentToLAG} we will define $\Omega$ as a PDF on the space of Lagrangian submanifolds. 

\item {\bf PDF on the space of Legendrian submanifolds.} Suppose that we can construct an ${\bf R}^{0|1}$-bundle over $M$, which we will call
   $\widehat{M}$, with a connection whose curvature is $\omega$. Lagrangian submanifolds on $M$ lift to
   Legendrian submanifolds in $\widehat{M}$. We can define $\Omega$ as a closed PDF on the space $\rm LEG$
   of Legendrian submanifolds in $\widehat{M}$. \marginpar{$\rm LEG$}

\item {\bf PDF on an equivalence class of actions.}
In BV formalism the choice of a Lagrangian submanifold $L\subset M$ is closely related to the 
choice of a quantization scheme. In other words, it is essentially the choice of a 
representative in a class of physically equivalent theories. Given $S_{\rm BV}$ and $L\in {\rm LAG}$, the 
restriction $\left.(S_{\rm BV})\right|_L$ gives a physical action functional which we use in the path integral. A 
different choice of $L$ gives a BRST equivalent action functional.
Therefore it would be natural to try to interpret $\Omega$ as a PDF on such an equivalence class. 
This, however, is not straightforward. The space of Lagrangian submanifolds is actually larger 
than the space of action functionals; to descend to the space of action functionals one has to 
take the factorspace over the symmetries of $S_{\rm BV}$. Generally speaking, $\Omega$ does not descend 
to this factorspace. We will study these issues in Section \ref{sec:EquivariantOmega}.
\end{enumerate}

\subsection{Definition of $\Omega$ as a PDF on $\widehat{G}$}\label{sec:OmegaOnPiTGxLAG}

For any function\footnote{we will actually use just $E(z) = e^z$} $E\;:\;{\bf C}\to {\bf C}$, consider the following pseudo-differential form on $\widehat{G}$:\marginpar{$E$}
\begin{align}
\Omega^{\{E\}} \in\; & \mbox{Fun}\left( (\Pi T\widehat{G})\times {\rm LAG} \right)
\\
\Omega^{\{E\}} =\; & \int_{\widehat{g}L} E({d\widehat{g}\widehat{g}^{-1}})\;\;\rho_{1\over 2}\label{OmegaM}
\end{align}
where $d\widehat{g}\widehat{g}^{-1}$ is the moment map of Section \ref{sec:MomentMap}; its main property is that for any function 
$F\in \mbox{Fun}(M)$:
\begin{equation} d_{(g)}(F\circ g) = \{d\widehat{g}\widehat{g}^{-1},F\,\}\circ g \end{equation}

\subsection{$\Omega$ is closed}

Under the assumption that $\Delta_{\rm can}\rho_{1\over 2}=0$ the form $\Omega$ is closed.

\paragraph     {Preparation for the proof of closedness}
Notice that for any even ${\cal H}\in \mbox{Fun}(M)$:\marginpar{$\cal H$}
\begin{align}
{\cal H} \Delta_{\rm can}(E({\cal H})\rho_{1\over 2}) - \Delta_{\rm can}({\cal H}E({\cal H})\rho_{1\over 2}) + {1\over 2} \{{\cal H},{\cal H}\}E'({\cal H})\rho_{1\over 2} \;+\; & \label{BeforeIntegration}
\\
+\;\Delta_{\rm can}\left(\left(\int d{\cal H}\,E({\cal H})\right)\rho_{1\over 2}\right) - \left(\int d{\cal H}\,E({\cal H})\right)\Delta_{\rm can}\rho_{1\over 2} & \;=\;0
\end{align}
We can interpret this formula using the notion of Lie derivative of half-density as follows:
\begin{equation}\label{InterpretationUsingLieDerivative} {\cal L}_{\{f({\cal H}),\,\_\}} \rho_{1\over 2} = {\cal L}_{\{{\cal H},\,\_\}} \left(f'({\cal H})\rho_{1\over 2}\right) - {1\over 2} \{{\cal H},{\cal H}\}f''({\cal H})\rho_{1\over 2} \end{equation}
where $f({\cal H}) = \int d{\cal H} E({\cal H})$
(but we prefer to work with $E({\cal H})$ instead of $f({\cal H})$, because it is $E({\cal H})$ that enters in 
Eq. (\ref{OmegaM})). When ${\cal H}$ is  the moment map ${\cal H} = d\widehat{g}\widehat{g}^{-1}$, this can be combined with the
Maurer-Cartan Eq. (\ref{MaurerCartan}):
\begin{equation}
\left( d + {\cal L}_{\{d\widehat{g}\widehat{g}^{-1},\_\}}\right)
   \left(f'(d\widehat{g}\widehat{g}^{-1})\rho_{1\over 2}\right)\;=\;
   {\cal L}_{\{f(d\widehat{g}\widehat{g}^{-1}),\_\}}\rho_{1\over 2}
\end{equation}
(where $d$ only acts on $\widehat{g}$).

\paragraph     {Proof of closedness}
Taking $\cal H$ to be the moment map ${\cal H} = d\widehat{g}\widehat{g}^{-1}$, we get:
\begin{align}
& d\int_{gL}  E(d\widehat{g}\widehat{g}^{-1})\;\rho_{1\over 2} \; = \;
\\   
\;=\; & \phantom{d}\int_{gL}\;   d(E(d\widehat{g}\widehat{g}^{-1}))\,\rho_{1\over 2} \;-\; d\widehat{g}\widehat{g}^{-1}\;\Delta_{\rm can}\left(E( d\widehat{g}\widehat{g}^{-1})\;\rho_{1\over 2}\right)\;+\; \Delta_{\rm can}\left(d\widehat{g}\widehat{g}^{-1}\;E( d\widehat{g}\widehat{g}^{-1})\;\rho_{1\over 2}\right) \; =
\\  
=\; & \int_{gL}\;   
\begin{array}{l}
d(d\widehat{g}\widehat{g}^{-1})\;E'(d\widehat{g}\widehat{g}^{-1})\,\rho_{1\over 2} \;+\;
      {1\over 2}\{d\widehat{g}\widehat{g}^{-1},d\widehat{g}\widehat{g}^{-1}\}E'(d\widehat{g}\widehat{g}^{-1})\rho_{1\over 2}\;+
\cr
\; + \;
       \Delta_{\rm can}\left(
         \int d{\cal H}E({\cal H})\;\rho_{1\over 2}
       \right) - 
       \int d{\cal H}\,E({\cal H})\;\Delta_{\rm can}\rho_{1\over 2}
       \;
\end{array}
\\ 
=\; & 0 & \label{GenericM}
\end{align}
where we have taken into account the definition of the canonical odd Laplace operator and 
the fact that $\int_{gL}\Delta_{\rm can}(\ldots)=0$. We also used the Maurer-Cartan Eq. (\ref{MaurerCartan}).
In particular, let us take $E({\cal H})={\rm exp}({\cal H})$. Let us denote:  \marginpar{$\Omega\langle \_\rangle$}
\begin{align}\label{GeneralOmega}
 \Omega\langle F\,\rangle \;=\;&
 \int_{\widehat{g}L} \exp\left({d\widehat{g}\widehat{g}^{-1}}\right)\rho_{1\over 2} F
 \\  
 \Omega \;= \;&
 \Omega\langle 1 \rangle
\end{align}
Then Eq. (\ref{GenericM}) implies:\marginpar{$\Omega$}
\begin{equation}\label{OmegaIsIntertwiner}
 d\Omega\langle F\,\rangle \;= \;
   - \Omega\left\langle \rho_{1\over 2}^{-1}\Delta_{\rm can}(\rho_{1\over 2} F\,)\right\rangle
 = - \Omega\langle \Delta F\,\rangle  
\end{equation}
Also notice the following equation:
\begin{equation}\label{IotaOnOmega}
   \iota_{\{H,\_\}}\Omega\langle F\,\rangle = \Omega\langle  HF\,\rangle
\end{equation}

\subsection{Algebraic interpretation}

\subsubsection{Cone Lie superalgebra: general definition}
Let $G$ be a Lie group. It is possible to introduce the structure of a Lie group on $\Pi TG$. In fact:
\begin{equation}
\Pi T G = \mbox{Map}({\bf R}^{0|1},G)
\end{equation}
and the structure of the group is introduced by pointwise multiplication. Consider the 
corresponding Lie superalgebra:\marginpar{$\widetilde{\bf g}$}
\begin{equation}
\widetilde{\bf g} = \mbox{Lie}(\Pi TG)
\end{equation}
We can interpret $\widetilde{\bf g}$ as the algebra of maps from ${\bf R}^{0|1}$ to $\bf g$; therefore the elements of $\widetilde{\bf g}$ are $\bf g$-valued functions $f(\theta)$ of an odd parameter $\theta\in {\bf R}^{0|1}$. It is possible to extend $\widetilde{\bf g}$ by an extra odd element $\partial_{\theta}$ with the following commutation relations:
\begin{equation}
[\partial_{\theta},f(\theta)] = {\partial\over\partial\theta} f(\theta)
\end{equation}
We will call this extended algebra $\widetilde{\bf g}'$.\marginpar{$\widetilde{\bf g}'$}

\subsubsection{Cone Lie superalgebra associated to a BV algebra}
Let us consider a BV algebra $\cal G$ with the generator $\Delta$. Let ${\bf g}$ be the Lie superalgebra which 
is obtained from $\cal G$ by forgetting the associative algebra structure and flipping parity, 
and $\widetilde{\bf g}'$ the corresponding cone Lie superalgebra. We need to flip parity in order to turn 
$\{\_,\_\}$ into a Lie superalgebra operation. If the parity of $b$ as an element of $\cal G$, is $|b|$, 
then the parities of the corresponding elements of $\widetilde{\bf g}'$ are: $|(b,0)| = |b| + \bar{1}$ and 
$|(0,b)| = |b|$.

\refstepcounter{Theorems}
\paragraph     {Theorem \arabic{Theorems}:\label{theorem:ConeLieInBV}} 
The following formulas define the representation of $\widetilde{\bf g}'$ on $\cal G$:
\begin{align}
\rho(d) a = & \; - \Delta a
\\  
\rho((0, b)) a = & \; ba
\\  
\rho((b,0)) a = & \; [\rho((0,b)),\rho(d)] a\;=
\\  
= & \;(-)^{|b|}\Delta(ba) - b\Delta a
\end{align}
We have to check that:
\begin{align} 
[\rho((b,0)),\rho((0,c))]a \;=\;&
 \rho((b,0))\rho((0,c))a - (-)^{|c|(|b|+1)}\rho((0,c))\rho((b,0))a 
\nonumber\\   
\;=\;& \rho((0,\{b,c\}))a 
\end{align}
Indeed, we have:
\begin{align}
\rho((b,0))\rho((0,c))a \;=\; & (-)^{|b|}\Delta(bca) - b\Delta(ca) = (-)^{|b|}(\Delta b)ca + \{b,ca\} \;=\; 
\nonumber\\   
\;=\; & \{b,c\}a + (-)^{|b|}(\Delta b)ca + (-)^{|c|(|b|+1)}c\{b,a\}
\\[5pt]
\rho((0,c))\rho((b,0))a \;=\; & (-)^{|b|} c \Delta(ba) - cb \Delta a = (-)^{|b| + |c|(|b|+1)}(\Delta b)ca + c\{b,a\}
\end{align}

\subsubsection{Form $\Omega\langle\ldots\rangle$ as an intertwiner}\label{sec:OmegaIsIntertwiner}
Let us consider the particular case when $\cal G$ is the algebra of functions on the odd symplectic manifold $M$ (our BV phase space).

In this case, $\widetilde{\bf g}'$ naturally acts on the differential forms on $\widehat{G}$. Indeed, every element $\alpha\in {\bf g}=\mbox{Lie}(G)$ determines the corresponding right-invariant vector field on $\widehat{G}$. Then $(\alpha,0)$ would act as a Lie derivative along this vector field, and $(0,\alpha)$ acts as a contraction. 

We can consider $\Omega\langle\_\,\rangle$ as a linear map from ${\cal G}$ to the space of differential forms on $\widehat{G}$; for each $a\in{\cal G}$, this map computes $\Omega\langle a\rangle$ --- the corresponding differential form. Eqs. (\ref{OmegaIsIntertwiner}) and (\ref{IotaOnOmega}) can be interpreted as saying that $\Omega\langle\_\,\rangle$ is an intertwiner:
\begin{equation} x\Omega\langle a\rangle = \Omega\langle \rho(x)a \rangle \end{equation}
In the language of half-densities, we have an intertwiner between:
\begin{itemize}
\item action of $\widetilde{\bf g}'$ on half-densities in $M$:
\begin{align}
R(d)\rho_{1\over 2}\;=\; & -\Delta \rho_{1\over 2}
\\
R(\iota_{\xi_H})\rho_{1\over 2}\;=\; & H\rho_{1\over 2}
\\
R(\xi_H)\rho_{1\over 2}\;=\; & (-)^{\overline{H}} \Delta(H\rho_{1\over 2}) - H\Delta\rho_{1\over 2}
\end{align}
and
\item action of $\widetilde{\bf g}'$ on PDFs on $\widehat{G}$:
\begin{align}
R(d)\alpha\;=\; & d\alpha
\\
R(\iota_{\xi_H})\alpha\;=\; & \iota_{\xi_H}\alpha
\\
R(\xi_H)\alpha\;=\; & {\cal L}_{\xi_H}\alpha
\end{align}
\end{itemize}

\subsection{Descent to $\rm LAG$}\label{sec:DescentToLAG}
\subsubsection{Deviation of $\Omega$ from being horizontal}
In Section \ref{sec:OmegaOnPiTGxLAG}  we defined a form $\Omega$ on $\Pi T\widehat{G}\times {\rm LAG}$. The action of $\widehat{G}$ on $\rm LAG$ defines
a natural projection:
\begin{equation}
\widehat{\pi} \;:\; \Pi T\widehat{G}\times {\rm LAG}\rightarrow \Pi T{\rm LAG}
\end{equation}
It is natural to ask if $\Omega$ is constant along the fibers of $\widehat{\pi}$. The answer is, strictly speaking, no,
but the dependence on the fiber is rather mild.
Let us present $\Pi T{\rm LAG}$ as follows:
\begin{equation} \Pi T{\rm LAG} = (\Pi T \widehat{G})\times_{\widehat{G}}{\rm LAG} \end{equation}
where $\times_{\widehat{G}}$ means factor over the symmetry:
\begin{equation}\label{GtimesLAGSymmetry} (\widehat{g},L)\mapsto (\widehat{g}\widehat{f},\widehat{f}^{-1}L) \end{equation}
For the descent to work, $\Omega$ should be base, i.e. both invariant and horizontal, with respect to (\ref{GtimesLAGSymmetry}). Actually, it is invariant and almost horizontal. The invariance is straightforward. Let us study the question of horizontality, and understand why it is ``almost horizontal'' instead of ``horizontal''. Let $\xi \in \widehat{\bf g}$ denote a Hamiltonian whose flux preserves the Lagrangian submanifold $L$. The horizontality would be equivalent to the statement that $\iota_{{\rm Ad}(\widehat{g})\xi}\Omega$ is zero. But in fact, Eq. (\ref{IotaOnOmega}) implies:
\begin{equation} \iota_{{\rm Ad}(\widehat{g})\xi}\Omega = \int_{\widehat{g}L}({\rm Ad}(\widehat{g})\xi) e^{d\widehat{g}\widehat{g}^{-1}}\rho_{1\over 2} \end{equation}
Notice that $\xi\in \mbox{St}(L)$ implies ${\rm Ad}(\widehat{g})\xi\in \mbox{St}(\widehat{g}L)$. Apriori this only implies that $({\rm Ad}(\widehat{g})\xi)|_{\widehat{g}L}$ is a constant (but not necessarily zero).

This is potentially a problem (we certainly do want $\Omega$ to descend on $\rm LAG$). We will now outline some possible ways of resolving this problem.

\subsubsection{Use ghost number symmetry}
In string theory there is usually an action of $U(1)$  called ``ghost number symmetry''. 
Various objects are either $U(1)$-invariant or have definite charge.

Let us restrict ourselves to only considering those Lagrangian submanifolds which are invariant under 
this $U(1)$ ({\it i.e.} request that the orbits of the $U(1)$ be tangent to the Lagrangian submanifold) and
request that $d\widehat{g}\widehat{g}^{-1}$ have ghost number $-1$.
This eliminates the possibility of adding a constant to $d\widehat{g}\widehat{g}^{-1}$ and therefore renders $(d\widehat{g}\widehat{g}^{-1})_L$ 
unambigously defined from the variation of $L$. Our form $\Omega$ is now horizontal, and descends from $G$ to $\rm LAG$.

\subsubsection{Use transverse Lagrangian submanifold}
Suppose that we can find a Lagrangian submanifold $L^{\vee}\subset M$ which is transverse to all Lagrangian submanifolds from our family:
\begin{align}
(\widehat{g}L )\cap L^{\vee} \;\;\;=\; & p(\widehat{g}L) \quad \mbox{\tt\small (one point)}
\label{intersection-is-one-point}
\end{align}
where $p(\widehat{g}L)$ is a marked point on every $\widehat{g}L$.

Let ${\cal Z}_{L^{\vee}}\subset \mbox{Fun}(M)$ denote the subspace of those Hamiltonians which vanish on $L^{\vee}$. Since $L^{\vee}$ is chosen to be Lagrangian, $\Pi {\cal Z}_{L^{\vee}}$ is a Lie subalgebra of ${\bf g}=\Pi \mbox{Fun}(M)$. (Slightly smaller than the stabilizer of $L^{\vee}$.) It is intuitively clear (from counting the ``degrees of freedom'') that we can impose the following gauge condition: $\widehat{g}\in \exp(\Pi{\cal Z}_{L^{\vee}})\subset \mbox{St}(L^{\vee})\subset \widehat{G}$ This implies that $d\widehat{g}\widehat{g}^{-1}$ vanishes at the marked point: $(d\widehat{g}\widehat{g}^{-1})(p(\widehat{g}L)) = 0$

This eliminates the ambiguity of a constant in $d\widehat{g}\widehat{g}^{-1}$.

What happens if we change $L^{\vee}$ to another transversal Lagrangian submanifold $\tilde{L}^{\vee}$? Let us assume that we can choose:
\begin{align}
\widehat{f}\;\in\; & \mbox{St}(L)\subset \widehat{G}
\\
\mbox{\tt\small such that }  \widehat{f} \tilde{L}^{\vee} \;=\; & L^{\vee}
\end{align}
Then we just have to change:
\begin{equation} \widehat{g} \mapsto \widehat{g} \widehat{f} \end{equation}
As $\Omega$ is invariant, this would not change the result of the integration.

\subsubsection{Upgrade $\rm LAG$ to ${\rm LAG}_+$}
The most elegant solution is to use, instead of the space of Lagrangian submanifolds $\rm LAG$, 
the space ${\rm LAG}_+$ of Lagrangian submanifolds with marked point \cite{Mikhailov:2016myt}.

\subsection{Space of Legendrian submanifolds in $\widehat{M}$}

\subsubsection{Quantomorphisms}\label{sec:Quantomorphisms}
Suppose that there exists an  ${\bf R}^{0|1}$-bundle over $M$:
\begin{equation}
\widehat{M} \stackrel{\widehat{\pi}}{\longrightarrow} M
\end{equation}
with a connection such that the curvature is equal to $\omega$. Then we can realize the central extension $\widehat{G}$ 
as the group of automorphisms of this bundle.

We have the exact sequence: 
\begin{equation}
0\rightarrow \mbox{Inv}_{{\bf R}^{0|1}}(\Gamma(T\widehat{M}/M\,)) \rightarrow \mbox{Inv}_{{\bf R}^{0|1}}(\mbox{Vect}(\widehat{M}\,)) \xrightarrow{\widehat{\pi}_*} \mbox{Vect}(M\,)\rightarrow 0
\end{equation}
where $\mbox{Inv}_{{\bf R}^{0|1}}$ means that we should consider ${\bf R}^{0|1}$-invariant vector fields. The kernel of $\widehat{\pi}_*$
is the $0|1$-dimensional space $\mbox{Inv}_{{\bf R}^{0|1}}(\Gamma(T\widehat{M}/M\,))$. A connection is a split:
\begin{equation}
\underline{\rm lift}\;:\; \mbox{Vect}(M)\rightarrow \mbox{Inv}_{{\bf R}^{0|1}}(\mbox{Vect}(\widehat{M}))
\end{equation}
Suppose that we can find a ``symplectic potential'' $\alpha$ such that $\omega = d\alpha$. Then we can use it to 
construct the connection satisfying:
\begin{equation}
[\underline{\rm lift}(v_1),\underline{\rm lift}(v_2)]\;=\; \underline{\rm lift}([v_1,v_2]) + \omega(v_1,v_2) \partial_{\vartheta}
\end{equation}
where $\partial_{\vartheta}$ is the vector field arizing from the action of ${\bf R}^{0|1}$ on $\widehat{M}$. (We can think of $\vartheta$ as a coordinate 
in the fiber; it is only defined locally, but $\partial_{\vartheta}$ is globally well-defined.) Explicitly:
\begin{equation}\label{ConnectionAsLift}
\underline{\rm lift}(v) = v + (\iota_v\alpha)\partial_{\vartheta}
\end{equation}
Let us consider the subalgebra ${\bf g}\subset \mbox{Vect}(M)$ consisting of Hamiltonian vector fields. For every even (we will restrict to even vector fields for simplicity) $\{H,\_\}\in {\bf g}$ consider the following vector field on $\widehat{M}$:
\begin{equation}\label{DefHHat}
\hat{\xi}_H \;=\; \{H,\_\} + (\iota_{\{H,\_\}}\alpha + H) \partial_{\vartheta}
\end{equation}
It is defined to preserve the connection. An explicit calculation shows that the Lie derivative vanishes:
\begin{equation}
{\cal L}_{\hat{\xi}_H} (d\theta - \alpha) = 0
\end{equation}
Notice that the vertical component of $\hat{\xi}_H$ (with respect to the connection defined in Eq. (\ref{ConnectionAsLift})) is $H\partial_{\vartheta}$.
By construction, the space of vector fields of this form is closed under commutator. We can check it 
directly, using the formula: 
\begin{align}
& \left[\; \{H,\_\} + (\iota_{\{H,\_\}}\alpha + H)\partial_{\vartheta}\;,\; \{F,\_\} + (\iota_{\{F,\_\}}\alpha + F)\partial_{\vartheta} \;\right]\;=\;
\nonumber\\  
\;=\; & \{\{H,F\},\_\}\;+\; \left( {\cal L}_{\{H,\_\}}(\iota_{\{F,\_\}}\alpha) - {\cal L}_{\{F,\_\}}(\iota_{\{H,\_\}}\alpha) + 2\{H,F\} \right)\partial_{\vartheta}\;=\;
\nonumber\\    
\;=\; & \{\{H,F\}\,,\,\_\}\;+\; \left(\iota_{\{\{H,F\},\_\}}\alpha + \{H,F\}\right)\partial_{\theta}
\end{align}
As a Lie algebra this is $\Pi\mbox{Fun}(M)$. It integrates to the group of automorphisms of the 
fiber bundle $\hat{M}\to M$ which preserve the connection defined in Eq. (\ref{ConnectionAsLift}).

\subsubsection{Form $\Omega$ as a form on the space of Legendrian submanifolds $\rm LEG$}
A Legendrian submanifold in $\widehat{M}$ projects to a Lagrangian submanifold in $M$: 
\begin{equation}
\pi(\widehat{L}) = L
\end{equation}
This projection is typically {\em not} one-to-one; it is a cover. We can define $\Omega$
as a PDF on the space of Legendrian submanifolds by interpreting Eq. (\ref{GeneralOmega}) as integration over the projection:
\begin{equation}
\Omega(\widehat{g},d\widehat{g}) \;=\;\int_{\pi(\widehat{g}\widehat{L})}\exp\left(d\widehat{g}\widehat{g}^{-1}\right)\rho_{1\over 2} 
\end{equation}
This descends to a closed PDF on the space of Legendrian submanifolds $\rm LEG$.

\subsection{Reduction to integration over single Lagrangian submanifold}
Here we will explain that integration over a family of Lagrangian submanifolds can be reduced 
to integration over a single Lagrangian submanifold in some larger BV phase space.

A family of Lagrangian submanifolds $\{L(\lambda)\subset M|\lambda\in\Lambda\}$ defines a single Lagrangian 
submanifold in $L_{\Lambda} \;\subset\; M\times \Pi T^*(\Pi T\Lambda)$. We will now describe the construction of $L_{\Lambda}$.

As a first step, let us consider a submanifold $L''\subset M\times \Lambda$ which is defined as follows:
\begin{equation}
L'' = \{ (m,\lambda) \;|\; \lambda\in\Lambda\;,\;m\in L(\lambda) \}
\end{equation}
This can be promoted to a subspace $L'\subset M\times \Pi T\Lambda$: 
\begin{equation}
L' = \{ (m,\lambda,{[d\lambda]}) \;|\; (\lambda,{[d\lambda]})\in \Pi T\Lambda\;,\;m\in L(\lambda) \}
\end{equation}
Finally, we will construct $L_{\Lambda}$ as the following section of the vector bundle 
$M\times \Pi T^*(\Pi T\Lambda) \longrightarrow M\times \Pi T\Lambda$ restricted to $L'\subset M\times \Pi T\Lambda$:
\begin{align}
s\;:\;L' \rightarrow M\times \Pi T^*(\Pi T\Lambda) &
\\
s\,(m,\lambda,{[d\lambda]})\;=\;(m,\lambda,{[d\lambda]}\;,\;\; & \sigma(m) \;,\; 0)\label{nonzero-section}
\end{align}
where $\sigma(m)$ computes for every tangent vector to $\Lambda$ the value of its corresponding generating function on $m\in L(\lambda)$. This section defines our big Lagrangian submanifold:
\begin{equation} L_{\Lambda} = s(L') \end{equation}
There is a natural BV Hamiltonian $\widehat{d}$ on $\Pi T^*(\Pi T\Lambda)$. It descirbes the lift to $\Pi T^*$ of the 
natural nilpotent vector field $d$ on $\Pi T \Lambda$. We have:
\begin{align}
\int_{\Lambda}\Omega \;=\;\int_{L_{\Lambda}} \exp\left(S_{\rm BV} + \widehat{d}\;\right)
\\
\mbox{\tt\small where } \widehat{d} = {[d\lambda]}\lambda^{\star}
\end{align}
In the case of Yang-Mills theory, $\lambda$ is called $\bar{c}$ and $[d\lambda]$ is called $\pi$ (see Section \ref{sec:Faddeev-Popov}).

\section{Picture changing}
\subsection{Integration in two steps}
In our approach we integrate some closed form $\Omega$ over a cycle in the moduli space of Lagrangian submanifolds modulo gauge symmetries. The form $\Omega$ itself is defined in terms of the integration of $e^{S_{\rm BV}}$ over the Lagrangian submanifold (path integral). We therefore have a double integral: first the path integral, and then a finite-dimensional integral over a cycle in the moduli space.

However, we suspect that there is actually no fundamental distinction between these two steps. In principle, one can combine them into one integration.

It is often convenient to pull one or more integration out from the integration of $\Omega$ into the path integral or vice versa. This is ``picture changing''.

In the case of bosonic string we usually choose a Lagrangian submanifold corresponding to a fixed complex strucuture (or metric, depending on the flavour of the formalism). In this case the path integration contains the integration over the antifield to metric/complex structure, which is identified with the $b$-ghost. However, in principle we could also include into the path integral some partial integration over the moduli space of complex structures, and then integrate over the rest later.

It turns out that in the BV formalism, this (at least in some cases) corresponds to changing each Lagrangian submanifolds in the family into a different Lagrangian submanifold. This is a topologically nontrivial change, essentially a change in polarization.

\subsection{Baranov-Schwarz transform}\label{sec:BaranovSchwarz}
Any family of Lagrangian submanifolds can locally be considered as an orbit of some Lagrangian submanifold $L_0$
by an abelian subgroup of $G$: 
\begin{equation}
UL_0 = \{gL_0\;|\;g\in U\}
\end{equation}
where $U\subset G$ is an abelian subgroup. For every $L(u) = uL_0$ $u \in \in UL_0$, consider a submanifold 
$K(u)\subset L(u)$ which is the common zero set of all Hamiltonians of elements of $T_uU$ restricted 
on $L(u)$. Consider the union:
\begin{equation} \widetilde{L} = \bigcup\limits_{u\in U} K(u) \end{equation}
It is a Lagrangian submanifold in $M$. Moreover:
\begin{equation} \int_{\Pi T(UL_0)}\Omega = \int_{\widetilde{L}}\rho_{1\over 2} \end{equation}
Indeed, let us consider the $k$-form component of $\Omega$. We have to integrate it over some 
$k$-dimensional family of Lagrangian submanifolds. We parametrize the family by $s^1,\ldots,s^k$. 
Let us perform the Baranov-Schwarz transform by integrating over $d[ds]$. This turns $\Omega$ into 
an integral form. We get:
\begin{align}\label{InsertDeltaFunctions} 
& \int \prod_I ds^I \;d[ds^I]\; \int_{gL} \rho_{1\over 2} \exp\left(\sum_I ds^I{\partial g\over\partial s^I}g^{-1} \right) \;=\;
\\    
\;=\;& \int \prod_I ds^I \;\int_{gL} \rho_{1\over 2} \;\prod_{I=1}^k\delta\left( {\partial g\over\partial s^I}g^{-1} \right) 
\end{align}
(Remember that we identify ${\partial g\over\partial s^I}g^{-1}$ with the actual Hamiltonian, {\it i.e.} for us ${\partial g\over\partial s^I}g^{-1}$ is a 
function on $M$, and in particular a function on $L$.) This means that actually we are integrating 
not over the whole Lagrangian submanifold $L$, but over a submanifold $K\subset L$ of the codimension $k$, 
which is defined by the system of equations:
\begin{equation}
{\partial g\over\partial s^I}g^{-1} = 0\;, \quad I\in \{1,\ldots,k\}
\end{equation}
But on the other hand, remember that we have to integrate over the family of the dimension $k$. 
In other words, we lost $k$ integrations by inserting the $\delta$-functions (\ref{InsertDeltaFunctions}), 
but then regained them as integration over the family. This is equivalent to the ``90 degree rotation'' 
$L\mapsto \widetilde{L}$.

Therefore integration over a family of Lagrangian submanifolds is equivalent to the integration over 
a single "rotated" Lagrangian submanifold.

However, in the physical context of path integration, it is hard to make use of this equivalence.

\section{Equivariant $\Omega$}\label{sec:EquivariantOmega}
\subsection{Special canonical transformations}\label{sec:SpecialCanonicalTransformations}
A canonical transformation is called {\em special} if it preserves $\rho_{1\over 2}$. 
Let $SG$ denote the group of special canonical transformations, and $s{\bf g}$ its algebra.
We will also consider the corresponding centrally extended $S\widehat{G}$ and ${\bf s\widehat{g}}$. Elements of 
${\bf s}\widehat{\bf g}$ are those Hamiltonians which are annihilated by $\Delta_{\rho_{1/2}}$. 

Suppose that two Lagrangian submanifolds $L_1$ and $L_2$ are connected by a special
canonical transformation:
\begin{equation}
L_2 = g L_1\quad,\quad g\in SG
\end{equation}
Naively one would want to consider such Lagrangian submanifolds completely equivalent.
However, this does not work. The form $\Omega$ is not base with respect to the action
of $SG$ on $\rm LAG$. Moreover, at least in the finite-dimensional case the action of
$SG$ on $\rm LAG$ is usually transitive; the factor space $SG\backslash \rm LAG$  would be 
a discrete space. Therefore it does not make much sense to consider the factorspace over $SG$.

On the other hand, if we consider just $\rm LAG$ without any identifications, then there are 
no integration cycles. For our closed form $\Omega$ to produce invariants, we need some integration
cycles.

The solution is to consider a factorspace of $\rm LAG$ over a subgroup $H\subset SG$. In fact this 
$H$ can not be an arbitrary subgroup of $SG$; it has to be choosen carefully\footnote{Gauge 
symmetries were discussed in the context of BV formalism in \cite{Sen:1993ic,Grigoriev:1998gn};
the requirements for gauge symmetries which we need for our procedure appear to be more
restrictive.}.

\subsection{Straightforward descent does not work}
Let $\hat{h}$ be a special canonical transformation, i.e. $\hat{h}\in S\widehat{G}$. 
It follows immediately that $\Omega = \int_{gL} e^{d\widehat{g} \widehat{g}^{-1}}$ is invariant under the left shift $\widehat{g}\mapsto \widehat{h}\widehat{g}$. 
But $\Omega$ is not horizontal; for $\xi\in {\bf s\widehat{g}}$ we have:
\begin{equation} 
\iota_{\xi}\Omega = \int_{gL} \;\xi \;e^{ d\widehat{g}\widehat{g}^{-1}}\rho_{1\over 2} \neq 0 
\end{equation}
Therefore $\Omega$ does not descend from $\rm LAG$ to $SG\backslash {\rm LAG}$.

However, we will now identify a class of subalgebras ${\bf h}\subset {\bf sg}$ for which we can construct 
the base form $\Omega_{\bf h}^{\tt base}$. Being a base form, it descends to $H\backslash {\rm LAG}$, where $H$ is the Lie group 
generated by the flows of elements of $\bf h$.

\subsection{Equivariant half-densities}
Suppose that we have the following data:\marginpar{$H$}
\begin{enumerate}
\item A subgroup $H$ of the group $\widehat{G}$ (defined in Section \ref{sec:CanonicalTransformations}. 
\marginpar{$\bf h$}
Let $\bf h$ be the Lie algebra of $H$. For each $\xi\in {\bf h}$, let $\underline{\xi}\in\mbox{Fun}(M)$ denote the corresponding
Hamiltonian. \marginpar{$\underline{\xi}$}
\item A (nonlinear) map\footnote{Since we are discussing nonlinear maps, we need to think of $\bf h$ as
     a supermanifold, which happens to be a linear space. Suppose that, as a linear space,
     ${\bf h} = {\bf R}^{m|n}$; then $\xi \mapsto  \rho^{\tt C}_{1\over 2}(\xi)$ is a function of $m$
   even and $n$ odd variables.} from $\bf h$ to the space of half-densities on $M$: 
\begin{align}
{\bf h} \to & \mbox{ \tt\small half-densities on }M
\\   
\xi \mapsto &\; \rho^{\tt C}_{1\over 2}(\xi)
\end{align}
such that:\marginpar{$\rho_{1\over 2}^{\tt C}(\xi)$}
\begin{align}
  {\cal L}_{\xi}\rho^{\tt C}_{1\over 2}(\eta) \;=\;
  & 
    \left.{d\over dt}\right|_{t=0} \rho^{\tt C}_{1\over 2}(e^{t\,{\rm ad}_{\xi}}\eta)
    \label{RhoEquivariant}
  \\    
  \Delta_{\rm can} \left(\rho^{\tt C}_{1\over 2}(\xi)\right)\;=\;
  &
    \underline{\xi} \rho^{\tt C}_{1\over 2}(\xi)
    \label{RhoEquivariantlyClosed}
\end{align}
\end{enumerate}
Then the following equation is a cocycle of the $H$-equivariant Cartan complex on $\rm LEG$ (Section \ref{sec:OmegaSummary}):
\begin{equation}
\Omega(L,dL,\xi)  = \int_L e^{d\widehat{g}\widehat{g}^{-1}} \rho^{\tt C}_{1\over 2}(\xi)
\end{equation}
This is proven using Section \ref{sec:OmegaIsIntertwiner}. 

The compatibility of Eqs. (\ref{RhoEquivariant}) and (\ref{RhoEquivariantlyClosed}):
\begin{align}
  {\cal L}_{\xi} \rho^{\tt C}_{1\over 2}(\xi) \;=\;
  &  
    -\Delta_{\rm can}\left( \underline{\xi}\rho_{1\over 2}(\xi) \right)
    - \underline{\xi} \Delta_{\rm can}\rho_{1\over 2}(\xi)\;=\;
\nonumber\\   
\;=\; &  
- \Delta_{\rm can}^2\left(\rho^{\tt C}_{1\over 2}(\xi)\right)  - \underline{\xi}^2\rho_{1\over 2}(\xi) \;=\;0
\end{align}
Here $\underline{\xi}^2=0$ because $\underline{\xi}$ is a fermion. (But the equivariance condition (\ref{RhoEquivariant}) is stronger than just
${\cal L}_{\xi} \rho^{\tt C}_{1\over 2}(\xi) = 0$, because it must hold for two arbitrary elements $\xi$ and $\eta$ of $\bf h$.)

It seems that the main ingredient in this approach is the choice of a group $H$. 
(In the case of bosonic string this is the group of diffeomorphisms.) 
The rest of the formalism is built around $H$. We suspect that $\rho_{1\over 2}^{\tt C}(\xi)$ is
more or less unambiguously determined by the choice of $H$; even $\rho_{1\over 2}^{\tt C}(0)$ is
already unambigously determined. Indeed, we will now see that the constraints arizing from
Eqs. (\ref{RhoEquivariant}) and (\ref{RhoEquivariantlyClosed}) are very tight.

\subsection{Expansion of $\Omega(L,dL,\xi)$ in powers of $\xi$}
\subsubsection{Derivation of a special subspace ${\cal F}\subset \mbox{Fun}(M)$}
Let us expand $\rho^{\tt C}_{1\over 2}(\xi)$ in powers of $\xi$:\marginpar{$\Phi\langle\xi\rangle$}
\begin{align}\label{ExpansionOfRho}
\rho^{\tt C}_{1\over 2}(\xi) = \rho^{(0)}_{1\over 2}\left( 1 + \Phi\langle\xi\rangle + \ldots\right)
\end{align}

\commentstarts{\small
we use angular brackets to emphasize linear depnendence: $\Phi\langle\xi\rangle$ is some 
linear function of $\xi \in {\bf h}$}
\commentends

\vspace{10pt}
\noindent
Eq. (\ref{RhoEquivariantlyClosed}) implies that $\rho^{(0)}_{1\over 2}$ satisfies the Master Equation. 
In this Section we will use $\rho^{(0)}_{1\over 2}$ to define the odd Laplace operator on functions:
\begin{equation}
\Delta = \Delta_{\rho^{(0)}_{1\over 2}}
\end{equation}
\marginpar{$\Delta$ here}
Eq. (\ref{RhoEquivariant}) implies that $\rho^{(0)}_{1\over 2}$ is $H$-invariant; using Eq. (\ref{ViaLieDerivative}) we derive:
\begin{equation}
{\cal L}_{\{F,\_\}}\rho^{(0)}_{1\over 2}\;=\; 
(-)^{\overline{F}}\Delta_{\rm can}\left(F\rho_{1\over 2}\right)\;=\;0
\end{equation}
Therefore the Hamiltonians generating $\bf h$ should be all $\Delta$-closed. Moreover, Eq. (\ref{RhoEquivariantlyClosed}) implies:
\begin{equation}
\Delta \Phi\langle\xi\rangle = \xi
\end{equation}
so they are actually all $\Delta$-exact. Eq. (\ref{RhoEquivariant}) implies:
\begin{equation}\label{EquivarianceOfPhi}
\{\Delta\Phi\langle\xi\rangle \,,\,\Phi\langle\eta\rangle\}\;=\; \Phi\langle[\xi,\eta]\rangle
\end{equation}
The image of $\Phi$, as a linear statistics-reversing map ${\bf h}\to \mbox{Fun}(M)$, will be called $\cal F$.
\marginpar{$\cal F$}
Notice that the inverse map to $\Phi\;:\;{\bf h}\rightarrow {\cal F}$ is $\Delta\;:\;{\cal F}\rightarrow {\bf h}$. To summarize:
\begin{equation} 
{\bf h} \;=\; \Pi\Delta{\cal F} 
\end{equation}

\subsubsection{Properties of subspace ${\cal F}$}\label{sec:PropertiesOfF}
Therefore the existence of equivariant analogue of $\Omega$ implies the existence of 
a subspace ${\cal F}\subset \mbox{Fun}(M)$ satisfying some special properties, which we will now study.
Let us define the bracket:
\begin{equation}
\label{DefLieAlgeraOfDiffeomorphisms} [x,y] = \{x,\Delta y\} 
\end{equation}
which satisfies $\Delta [x,y] = \{\Delta x,\Delta y\}$. Eq. (\ref{EquivarianceOfPhi}) implies that:
\begin{align}
[x,y]\;\in\; & {\cal F}
\\    
\mbox{\tt\small and } [x,y] \;=\; & (-)^{\bar{x}\bar{y}+1}[y,x]
\end{align}
In other words:
\begin{itemize}
\item $\cal F$ is closed under $[\_\,,\_]$ and
\item $[\_\,,\_]$ is antisymmetric
\end{itemize}
Moreover:
\begin{itemize}
\item the Jacobi identity for $\{\_,\_\}$, implies the Jacobi identity for $[\_,\_]$; 
in other words $\Pi{\cal F}$ is a Lie superalgebra (isomorphic to $\bf h$)
\end{itemize}
It is enough to consider the case when $\cal F$ is even (all $\Phi$ are even). Then Eq. (\ref{EquivarianceOfPhi}) implies:
\begin{equation}
\Delta \{\Phi,\Phi\}\;=\;0
\end{equation}
where $\Phi = \Phi\langle\xi\rangle$ for some odd $\xi$. Moreover, Eq. (\ref{RhoEquivariantlyClosed}) at the second order of the
$\xi$-expansion implies that $\{\Phi,\Phi\}$ is $\Delta$-exact. In other words, exists a map\marginpar{$q$}
\begin{equation} 
q\;:\;{\cal F}\otimes {\cal F}\to \mbox{Fun}(M) 
\end{equation}
such that:
\begin{equation}\label{DefQ} 
\{x,y\} = \Delta q(x,y) 
\end{equation}

\subsubsection{Higher orders of $\xi$-expansion}
We can always write, without any further assumptions:\marginpar{$a(\xi)$}
\begin{align}\label{OmegaCartanDensity}
\rho^{\tt C}_{1\over 2}(\xi)\;=\; & e^{a(\xi)}\rho^0_{1\over 2}
\end{align}
So defined $a(\xi)$ should satisfy (we abbreviate $\Phi\langle\xi\rangle$ to just $\Phi$):
\begin{align}
\Delta a + {1\over 2} \{a,a\} \;=\;& \Delta\Phi 
\label{EquationForLogForEquivariant} 
\\    
\{\Delta \Phi\langle\eta\rangle \,,\,a(\xi)\}\;=\;&
\left.{d\over dt}\right|_{t=0} a(e^{t\,{\rm ad}_{\eta}}\xi)
\label{EquivarianceOfA}
\end{align}
The first two terms in the $\xi$-expansion (which is the same as $\Phi$-expansion) are:
\begin{equation} 
a = \Phi - {1\over 2} q(\Phi,\Phi) + \ldots 
\end{equation}
The consistency of Eq. (\ref{EquationForLogForEquivariant}) requires that $\Delta \{a,a\} =0$ at every order in $\Phi$. This
automatically implies:
\begin{equation} 
\Delta \{a,a\} = 2\{\Delta a,a\} = -\{\{a,a\},a\} + 2\{\Delta\Phi,a\} 
\end{equation}
The first term $-\{\{a,a\},a\}$ is automatically zero because of the Jacobi identity. The 
vanishing of $2\{\Delta\Phi,a\}$ follows from Eq. (\ref{EquivarianceOfA}):
\begin{equation} \{\Delta \Phi, a(\Phi)\} = a(\{\Delta\Phi,\Phi\}) = a (0) = 0 \end{equation}
Therefore $\Delta$-closedness of $\{a,a\}$ is automatically satisfied order by order. But
its $\Delta$-exactness is a nontrivial {\bf additional assumption} which has to be verified on
case-by-case basis. Moreover, $\{a,a\}$ should be $\Delta$-exact in an equivariant manner,
{\it i.e.} satisfying Eq. (\ref{EquivarianceOfA}). To the first order in $\Phi$:
\begin{align}
a\;=\;\Phi - {1\over 2}\Delta^{-1}\{ & \Phi - {1\over 2}\Delta^{-1}\{ & \Phi - {1\over 2}\Delta^{-1}\{\Phi,\Phi\} & , &
\nonumber\\
 & & \Phi - {1\over 2}\Delta^{-1}\{\Phi,\Phi\} & \} & ,
\nonumber\\
 & \Phi - {1\over 2}\Delta^{-1}\{ & \Phi - {1\over 2}\Delta^{-1}\{\Phi,\Phi\} & , &
\nonumber\\
 & & \Phi - {1\over 2}\Delta^{-1}\{\Phi,\Phi\} & \} & \} \;+\; o(\Phi^4)
\end{align}
The terms of the order $\Phi^3$ are:
\begin{equation}\label{DeltaInverseAtThirdOrder}
{1\over 2}\Delta^{-1}\{\Phi,\Delta^{-1}\{\Phi,\Phi\}\} \;=\; {1\over 2}\Delta^{-1}\{\Phi,q(\Phi,\Phi)\} 
\end{equation}
The terms of the order $\Phi^4$ are:
\begin{align}
     & -{1\over 2}\Delta^{-1}\left(\{\Phi,\Delta^{-1}\{\Phi,\Delta^{-1}\{\Phi,\Phi\}\}\} \,+\,{1\over 4}\{\Delta^{-1}\{\Phi,\Phi\},\Delta^{-1}\{\Phi,\Phi\}\}\right)
\end{align}
Notice that:
\begin{align}
     & \Delta\left(\{\Phi,\Delta^{-1}\{\Phi,\Delta^{-1}\{\Phi,\Phi\}\}\} + {1\over 4}\{\Delta^{-1}\{\Phi,\Phi\},\Delta^{-1}\{\Phi,\Phi\}\}\right)\;=
    \\
    = & \{\Delta\Phi,\Delta^{-1}\{\Phi,\Delta^{-1}\{\Phi,\Phi\}\}\}
\end{align}
This is zero by our assumption that the function $\Phi\mapsto \Delta^{-1}\{\Phi,\Delta^{-1}\{\Phi,\Phi\}\}$ is 
$\bf h$-invariant. This means that in Eq. (\ref{DeltaInverseAtThirdOrder}) we are taking $\Delta^{-1}$ of a $\Delta$-closed expression. 
We must assume that this expression is also $\Delta$-exact, so we could take $\Delta^{-1}$ of it. 
This assumption has to be verified order by order.

\subsection{{\bf Summary:} constraints on gauge symmetry}\label{sec:SummaryOfConditions}
The symmetry group $H$ with Lie algebra $\bf h$ acts on the odd symplectic manifold $M$ (the BV phase
space), in such a special way that the following requirements are satisfied.

\vspace{10pt}\noindent
We require to exist the solution of Eqs. (\ref{RhoEquivariant}) and (\ref{RhoEquivariantlyClosed}) of the form of Eq. (\ref{OmegaCartanDensity}). 
We require that $a(\xi)$ be smooth at $\xi=0$. This implies that $\rho^{(0)}_{1\over 2}$ defined in Eq. (\ref{ExpansionOfRho}) 
satisfies the Quantum Master Equation: $\Delta_{\rm can}\rho^{(0)}_{1\over 2}=0$ and moreover that $\rho^{(0)}_{1\over 2}$ is $\bf h$-invariant. 
This means that every generator of $\xi\in \bf h$ is $\Delta$-closed, where $\Delta=\Delta_{\rho^{(0)}_{1\over 2}}$ is the odd Laplace 
operator on functions constructed from $\rho^{(0)}_{1\over 2}$. Moreover, we assume that every element of $\xi \in \bf h$ 
is actually $\Delta$-exact, {\it i.e.} exists $\Phi\langle\xi\rangle$ such that:
\begin{equation}
\xi = \Delta\Phi\langle\xi\rangle
\end{equation}
The linear subspace $\cal F$ of $\mbox{Fun}(M)$  generated by $\Phi\langle\xi\rangle$, $\xi\in {\bf h}$,
should satisfy the properties described in Section \ref{sec:PropertiesOfF}. (In simplest cases all elements of $\cal F$ 
are in involution with each other.) Moreover, we must require the existence of a solution $a$ to 
Eqs. (\ref{EquationForLogForEquivariant}) and (\ref{EquivarianceOfA}).

\subsection{Deformations of equivariant half-density}
What can we say about the moduli space of solutions of Eqs. (\ref{RhoEquivariant}), (\ref{RhoEquivariantlyClosed})? Suppose that we
can find another solution $\tilde{\rho}^{\tt C}_{1\over 2}(\xi)$:
\begin{align}
\tilde{\rho}^{\tt C}_{1\over 2}(\xi)\;=\; & R(\xi)\rho^{\tt C}_{1\over 2}(\xi)
\\      
R\;\in\;&\mbox{Fun}({\bf h}\times M)
\end{align}
Eqs. (\ref{RhoEquivariant}), (\ref{RhoEquivariantlyClosed}) imply that $R$ should satisfy:
\begin{align}
{\cal L}_{\xi}R(\eta) \;=\;& \left.{d\over dt}\right|_{t=0} R(e^{t\,{\rm ad}_{\xi}}\eta)
\label{RIsEquivariant}\\   
\Delta_{\rho^{\tt C}_{1\over 2}(\xi)} \, R(\xi) \;=\; & 0
\label{DeltaRIsZero}
\end{align}
This means that infinitesimal deformations of $\rho^{\tt C}_{1\over 2}$ are determined by the cohomology of $\Delta_{\rho^{\tt C}_{1\over 2}(\xi)}$
in the space of $\bf h$-invariant functions ${\bf h}\to \mbox{Fun}(M)$. 

Let us assume that $R(\xi)$ is a smooth function of $\xi$ and can be expanded in series in $\xi$.
Suppose that the leading term is of the order $n$ in $\xi$ (in other words, $R$ has  a zero
of the order $n$ at $\xi=0$):
\begin{align}
R(\xi) \;=\; R_n\langle\xi^{\otimes n}\rangle + R_{n+1}\langle\xi^{\otimes n+1}\rangle + \ldots
\end{align}
Eq. (\ref{DeltaRIsZero}) implies:
\begin{equation}
\Delta R_n \;=\;0
\end{equation}
where $\Delta = \Delta_{\rho_{1\over 2}(0)} = \Delta^{(0)} + \{S_{\rm BV}\,,\,\_\}$. Therefore we can think of $R_n$ as a map:
\begin{equation}\label{GhostNumberOfR}
S^n{\bf h} \stackrel{R_n}{\longrightarrow} 
\left[\mbox{ \tt\small integrated vertex operators of ghost number }-2n\,\right]
\end{equation}
commuting with the action of $\bf h$. This means that the space of deformations can be computed
by the spectral sequence whose initial page is the cohomology of $\Delta$ in the space
of invariant polynomials on $\bf h$ with values in functions on $M$; the ghost number should
correlate with the degree of polynomial as in Eq. (\ref{GhostNumberOfR}).

\subsection{Base form}\label{sec:BaseForm}
Let $\Omega_{\bf h}^{\tt C}$ denote the equivariant form, given by Eq. (\ref{OmegaCartanDensity}). We will denote the corresponding base form $\Omega^{\tt base}_{\bf h}$: 
\begin{equation}
\Omega^{\tt base}_{\bf h} \in \Omega^{\bullet}({\rm LAG})\quad\mbox{ \tt\small base for } \begin{array}{c}{\rm LAG}\cr \cr \downarrow\cr \cr H \backslash {\rm LAG}\end{array}
\end{equation}
We will work under the assumption that the action of $\bf h$ on $\rm LAG$ does not have fixed points;  
${\rm LAG}\rightarrow H\backslash{\rm LAG}$ can be considered a principal $H$-bundle. 
In order to construct the base form $\Omega^{\tt base}_{\bf h}$ from the Cartan's $\Omega^{\tt C}_{\bf h}$, we first choose on this principal bundle some connection ${\cal A}$. (We understand the connection as a $\cal F$-valued 1-form on $\rm LAG$ computing the ``vertical component'' of a vector.) Then we apply the horizontal projection i.e. replace
\begin{align}
d\widehat{g} \widehat{g}^{-1} \; & \mapsto d\widehat{g} \widehat{g}^{-1} - {\cal A}
\\
\mbox{\tt\small in } \Omega^{\tt C}_{\bf h} \; & = \int_{gL} e^{d\widehat{g} \widehat{g}^{-1}}\Gamma(\Phi)\rho_{1\over 2}
\end{align}
Finally, we replace $\Phi$ with the curvature $d{\cal A} + {\cal A}^2$ of the connection $\cal A$; we get:
\begin{equation}
\Omega^{\tt base}_{\bf h} = \int_{gL} e^{d\widehat{g} \widehat{g}^{-1} - {\cal A}} \Gamma( d{\cal A} + {\cal A}^2 )\rho_{1\over 2}
\end{equation}
\subsection{Example: Duistermaat-Heckman integration}
Consider the case when our odd symplectic submanifold is the odd cotangent bundle of some symplectic supermanifold $X$:
\begin{equation} M = \Pi T^*X \end{equation}
Let $p\;:\; M\to X$ be the projection. Let $\pi$ denote the Poisson bivector of $X$. It defines the following solution of the Quantum Master Equation:
\begin{align}
\rho_{1\over 2}(x,x^{\star},{\bf B}) \;=\; & \exp(\pi^{ij}(x)x^{\star}_ix^{\star}_j)
\\
\mbox{\tt\small where } & {\bf B}\;=\;\{e_1,\ldots,e_n,f^1,\ldots,f^n,\,\epsilon^{\vee 1},\ldots,\epsilon^{\vee n}, \eta_{\vee 1},\ldots, \eta_{\vee n}\}
\\  
\mbox{\tt\small where } &
\{e_1,\ldots,e_n,f^1,\ldots,f^n\}
\mbox{ \tt\small is a Darboux basis in } T_x X
\\   
\mbox{\tt\small and } & 
\{\epsilon^{\vee 1},\ldots,\epsilon^{\vee n}, \eta_{\vee 1},\ldots, \eta_{\vee n}\} 
\mbox{ \tt\small its dual}
\end{align}
\commentstarts{\small
In this formula we define a half-density at $m=(x,x^{\star})\in M$ as a function on the space of bases (``tetrads'') in $T_m M \simeq T_x X\oplus \Pi T^*_xX$, such that the value on two different bases differs by a multiplication by the super-determinant.
}\commentends

Let $H$ be a subgroup of the group of canonical transformations of $X$. A canonical transformation $h$ of $X$ can be lifted to a BV-canonical transformation $\Pi T^* h$ of $\Pi T^* X$ as follows:
\begin{equation}\label{lift-of-canonical-transformation} 
(\Pi T^* h)(\xi,x) = ((h^{-1}(x)_*)^*\xi,h(x))
\end{equation}
An infinitesimal canonical transformation of $X$ generated by a Hamiltonian $F$ lifts to an infinitesimal BV-canonical transformation of $\Pi T^* X$ generated by the BV-Hamiltonian $\Delta_{\rho_{1\over 2}} (p^*F)$.
Every point $x\in X$ defines a Lagrangian submanifold in $\Pi T^* X$: the fiber $\Pi T^*_x X$. Therefore, we can think of $X$ as a family of Lagrangian submanifolds in $\Pi T^* X$.
In this case $\cal F$ is the $p^*$ of the subspace of Hamiltonians generating $H$. The equivariant form $\Omega$ becomes:
\begin{align}
\phantom{\int_{\Pi T^*_xX}}\Omega(x,[dx],F) \;=\; & \int_{\Pi T^*_xX} \rho_{1\over 2} \exp\left([dx]^i x^{\star}_i + F(x)\right)\;=
\\
\;=\; & \exp\left(\varpi_{ij}(x)[dx]^i [dx]^j + F(x)\right)
\end{align}
where $\varpi = \pi^{-1}$.

\subsection{Example: an interpretation of Cartan complex}
Let us consider the case:
\begin{equation} M = \Pi T^* (\Pi T X) \end{equation}
The odd tangent bundle $\Pi T X$ has a natural volume form. This induces a half-density on $M$.
Let $\rho^{(0)}_{1\over 2}$ be the half-density on M induced from the volume form on $\Pi TX$. It satisfies the 
Quantum Master Equation.
Now consider a different half-density, which also satisfies the Quantum Master Equation\footnote{We can replace $e^{\widehat{d}}$ with an arbitrary function of $\widehat{d}$. This will still satisfy the Quantum Master Equation.}:
\begin{equation} \rho_{1\over 2}\;=\; e^{\widehat{d}} \rho^{(0)}_{1\over 2} \end{equation}
where $\widehat{d}$ is the BV Hamiltonian generating the lift to $\Pi T^*(\Pi TX)$ of the canonical odd vector field $d$ on $\Pi TX$.

Suppose that a Lie group $H$ acts on $X$. This action can be lifted to $\Pi TX$ and $\Pi T^*(\Pi TX)$. An infinitesimal action of $\xi \in {\bf h} = \mbox{Lie}(H)$ on $M$ is generated by the BV Hamiltonian:
\begin{equation} \widehat{{\cal L}_{\xi}} = \{\widehat{d},\widehat{\iota_{\xi}}\} \end{equation}
Therefore, in this case:
\begin{align}
a(\xi) = & \Phi\langle\xi\rangle = \widehat{\iota_{\xi}}
\\
\rho^{\tt C}_{1\over 2}(\xi) = & e^{\widehat{d} + \widehat{\iota_{\xi}}} \rho^{(0)}_{1\over 2}
\end{align}
It satisfies Eq. (\ref{RhoEquivariantlyClosed}):
\begin{equation}
\Delta \left(e^{\widehat{d} + \widehat{\iota_{\xi}}} \rho^{(0)}_{1\over 2}\right) \;=\; \widehat{{\cal L}_{\xi}}e^{\widehat{d} + \widehat{\iota_{\xi}}} \rho^{(0)}_{1\over 2}
\end{equation}
This formula can be generalized as follows. Suppose that $\alpha$ is a function on ${\bf h}\times \Pi TX$. For any $\xi\in {\bf h}$, we think of $\alpha(\xi)$ as a function on $\Pi TX$, i.e. a pseudo-differential form on $X$. With a slight abuse of notaions, $\alpha(\xi)$ will also denote the pullback of $\alpha(\xi)$ along the projection, from $\Pi TX$ to $\Pi T^*(\Pi TX)$. Then we have:
\begin{align}
\Delta \left(\alpha(\xi) e^{\widehat{d} + \widehat{\iota_{\xi}}} \rho^{(0)}_{1\over 2}\right) \;=\; & \widehat{{\cal L}_{\xi}}\alpha(\xi) e^{\widehat{d} + \widehat{\iota_{\xi}}} \rho^{(0)}_{1\over 2} \;+
\\
& + ((d+\iota_{\xi})\alpha(\xi))\,e^{\widehat{d} + \widehat{\iota_{\xi}}} \rho^{(0)}_{1\over 2}
\end{align}
Therefore $\alpha\;\mapsto\;\alpha e^{\widehat{d} + \widehat{\iota}}$ is an intertwiner from the Cartan complex to the complex of equivariant half-densities on $\Pi T^*(\Pi TX)$.

\subsection{Equivariant effective action}
When some fields are integrated out, we get an ``effective action'' for the remaining fields.

\paragraph     {Mini-review of BV effective action}
Let $(M,\omega)$ be an odd symplectic supermanifold (the BV phase space), and $\rho_{1\over 2}$ be a half-density
on $M$.
Suppose that we are given a submanifold $E\subset M$ which comes with the structure of a fiber 
bundle with fibers isotropic submanifolds over some base $B$. (This can be thought of as a family 
of isotropic submanifolds.) For every $b\in B$ we have the corresponding fiber, an isotropic 
submanifold which we denote $I(b)$. Moreover, we require:
\begin{equation} \mbox{ker} \,\omega|_{E} = TI \end{equation}
(i.e. the degenerate subspace of the restriction of $\omega$ to $E$ is the tangent space of the fiber). 
Since $d\omega = 0$, this condition implies that the Lie derivative of $\omega|_E$ along the fiber vanishes,
and therefore $\omega$ defines an odd symplectic form on the base which we will denote $\omega_B$.

Finally, we require certain maximality property:
\begin{align}
&\mbox{\tt\small E cannot be embedded in any larger} 
\nonumber\\   
&\mbox{\tt\small submanifold of $M$ where $TI$ would still be isotropic}\label{maximality-requirement}
\end{align}
Let $L\subset B$ be a Lagrangian submanifold of $B$. Under the above conditions, it can be lifted to 
a Lagrangian submanifold in $M$ as $\pi^{-1} L$, where $\pi$ is a natural projection:
\begin{equation} \pi\;:\; E\to B \end{equation}
We then define a half-density $\pi_*\rho_{1\over 2}$ on $B$ so that for every Lagrangian submanifold $L\subset B$ and
every function $f\in \mbox{Fun}(B)$:
\begin{equation} \int_Lf\pi_*\rho_{1\over 2} = \int_{\pi^{-1}(L)} (f\circ \pi)\rho_{1\over 2} \end{equation}
The proof of existence of such $\pi_*\rho_{1\over 2}$ is similar to the proof of Theorem \ref{theorem:DeltaCanonical}. We define a measure
$\mu_L[\rho_{1\over 2}]$ on $L$ as push-forward of $\left.\rho_{1\over 2}\right|_{\pi^{-1}L}$to $L$. We observe that for any canonical
transformation $g$ bringing fibers of $E$ into fibers: $g^*\mu_{gL}[\rho_{1\over 2}]=\mu_L[g^*\rho_{1\over 2}]$. Indeed:
\begin{align}
& \int_L (f\circ g) g^*\mu_{gL}[\rho_{1\over 2}]=\int_{gL} f \mu_{gL}[\rho_{1\over 2}]=
\int_{\pi^{-1}(gL)}(f\circ \pi)\rho_{1\over 2}=
\nonumber\\   
=\;&\int_{L}(f\circ g \circ \pi)g^*\rho_{1\over 2}
\end{align}
Then we define a half-density $\sigma_{1\over 2}[L,\rho_{1\over 2}]$ on $B$ so that $\mu_L[\rho_{1\over 2}]$ is its restriction to $L$. 
The same argument as in Appendix \ref{sec:ProofOfTheoremDef} shows that $\sigma_{1\over 2}[L,\rho_{1\over 2}]$ does not depend on $L$. We define
$\pi_*\rho_{1\over 2}= \sigma_{1\over 2}[L,\rho_{1\over 2}]$.

Suppose that we are given a function $\Psi$ on $M$ whose restriction on $E$ is constant along $I$. Then it defines a function on $B$ which we denote $\pi_*\Psi$:
\begin{equation} (\pi_*\Psi)\circ \pi = \Psi|_E \end{equation}
Notice that in this case the flux $\{\Psi,\_\}$ is tangent to $E$.

\commentstarts{\small
It is enough to prove that for any $\phi\in \mbox{Fun}(M)$ constant on $E$: $\{\Psi,\,\phi\,\} = 0$. From the maximality of $B$, as defined in Eq. (\ref{maximality-requirement}), follows that $\{\phi,\,\_\,\}$ is tangent to $I$. Since $\Psi$ is constant along $I$, it follows that indeed $\{\Psi,\,\phi\,\} = 0$.
}\commentends

\noindent
We have:
\begin{equation}\label{flow-commutes-with-projection} \forall m\in E\;:\; \pi(m)_* \left(\{\Psi,\,\_\,\}(m)\right) \;=\; \{\pi_*\Psi,\,\_\,\}(\pi(m)) \end{equation}
Equivalently, for two functions $\Psi_1$ and $\Psi_2$ on $M$ both constant along $I$:
\begin{equation} \{\Psi_1,\Psi_2\} = \{\pi_*\Psi_1,\pi_*\Psi_2\}\circ\pi \end{equation}
\commentstarts{\small
    In order to prove Eq. (\ref{flow-commutes-with-projection}), notice that any vector field tangent to $I$ can be written as $\sum_a f_a \{\phi_a,\_\}$ where $f_a$ are some functions on $M$ and $\phi_a$ are some functions on $M$ constant on $E$. The commutator with $\{\Psi,\,\_\,\}$ of such a vector field is $\sum_a \{\Psi,f_a\} \{\phi_a,\_\}$ — again tangent to $I$. This means that the flow of $\Psi$ brings fibers to fibers, which is equivalent to Eq. (\ref{flow-commutes-with-projection}).
}\commentends

We will now prove that for any $\Psi\in \mbox{Fun}(M)$ whose restriction on $E$ is constant along $I$:
\begin{equation}\label{push-forward-of-Lie-derivative} \pi_*\left({\cal L}_{\{\Psi,\_\}}\rho_{1\over 2}\right) \;=\; {\cal L}_{\{\pi_*\Psi,\_\}}\pi_*\rho_{1\over 2} \end{equation}
Indeed, for any ``test function'' $f\in \mbox{Fun}(B)$:
\begin{align}
\phantom{=\;}\delta_{\{\pi_*\Psi,\_\}}\int_L f\pi_*\rho_{1\over 2} \;=\;\int_L \left( \{\pi_*\Psi,f\} \pi_*\rho_{1\over 2} + (-)^{(\bar{\Psi}+1)\bar{f}}f{\cal L}_{\{\pi_*\Psi,\_\}}\pi_*\rho_{1\over 2} \right)\;=\;\label{first-line}
\\
\;=\;\delta_{\{\Psi,\_\}}\int_{\pi^{-1}L} (f\circ \pi)\rho_{1\over 2} \;=\; \int_{\pi^{-1}L}\left( \{\Psi,f\circ\pi\}\rho_{1\over 2} + (-)^{(\bar{\Psi}+1)\bar{f}}f{\cal L}_{\{\Psi,\_\}}\rho_{1\over 2} \right)\;=\;
\\
\;=\;\int_L \left(\{\pi_*\Psi,f\} \pi_*\rho_{1\over 2} + (-)^{(\bar{\Psi}+1)\bar{f}}f\pi_*\left({\cal L}_{\{\Psi,\_\}}\rho_{1\over 2}\right)\right)\label{third-line}
\end{align}
Equality of Lines (\ref{first-line}) and (\ref{third-line}) implies Eq. (\ref{push-forward-of-Lie-derivative}). In particular:
if $\rho_{1\over 2}$ satisfies the Quantum Master Equation on $M$, then $\pi_*\rho_{1\over 2}$ satisfies the Quantum Master Equation on $B$
Indeed, $\rho_{1\over 2}$ satisfying the Quantum Master Equation is equivalent to the statement that for any functions $\Psi$ and $F$:
\begin{equation}\label{QME-via-Lie-derivative} \delta_{\{\Psi,\_\}}\int_L F\rho_{1\over 2} = (-)^{\bar{F}+1} \int_L \Psi {\cal L}_{\{F,\_\}}\rho_{1\over 2} \end{equation}
When $\rho_{1\over 2}$ satisfies the QME on $M$, considering Eq. (\ref{QME-via-Lie-derivative}) with both $\Psi$ and $F$ constant along the fiber of $E\rightarrow B$ and using Eq. (\ref{push-forward-of-Lie-derivative}) proves that $\pi_*\rho_{1\over 2}$ also satisfies the QME.

\paragraph     {Equivariant case}
Now we will consider partial integration when the half-density satisfies the equivariant Master Equation. We show that the resulting effective theory still satisfies the equivariant Master Equation.

Suppose that a Lie group $H$ acts on $M$ by canonical transformations, and moreover that every Hamiltonian $\xi\in {\bf h}$ is constant on the fibers of $E\stackrel{\pi}\rightarrow B$:
\begin{equation} \forall \xi\in{\bf h} \; \exists \,\pi_*\xi\in \mbox{Fun}(B)\;:\; \xi|_E = (\pi_*\xi)\circ\pi \end{equation}
Suppose that we are given an equivariant half-density $\xi\mapsto \rho^{\tt C}_{1\over 2}(\xi)$ on $M$, i.e.:
\begin{align}
{\cal L}_{\eta}\rho^{\tt C}_{1\over 2}(\xi) \;=\; & \left.{d\over dt}\right|_{t=0} \rho^{\tt C}_{1\over 2}(e^{t\{\eta,\_\}}\xi)
\\
\Delta \rho^{\tt C}_{1\over 2}(\xi) \;=\; & \xi \rho^{\tt C}_{1\over 2}(\xi)\label{equivariant-QME}
\end{align}
Then the result of the partial integration is an equivariant half-density on $B$:
\begin{align}
{\cal L}_{\eta}\pi_*\rho^{\tt C}_{1\over 2}(\xi) \;=\; & \left.{d\over dt}\right|_{t=0} \pi_*\rho^{\tt C}_{1\over 2}(e^{t\{\eta,\_\}}\xi)
\\
\Delta \pi_*\rho^{\tt C}_{1\over 2}(\xi) \;=\; & (\pi_*\xi) \pi_*\rho^{\tt C}_{1\over 2}(\xi)\label{equivariant-QME-on-base}
\end{align}
Indeed, Eq. (\ref{equivariant-QME}) is equivalent to the statement that for any two functions $\Psi$ and $F$:
\begin{equation}\label{equivariant-QME-via-Lie-derivative} \delta_{\{\Psi,\_\}}\int_L F\rho_{1\over 2}(\xi) = (-)^{\bar{F}+1} \int_L \left( \Psi {\cal L}_{\{F,\_\}}\rho_{1\over 2}(\xi) + \Psi F \xi \rho_{1\over 2}(\xi)\right) \end{equation}
Eq. (\ref{equivariant-QME-on-base}) follows from Eqs. (\ref{push-forward-of-Lie-derivative}) and (\ref{equivariant-QME-via-Lie-derivative}) considering the case when both $\Psi$ and $F$ are 
constant along the fiber of $E\rightarrow B$. 

In this sense, the property of solving the equivariant Master Equation (\ref{RhoEquivariant}), (\ref{RhoEquivariantlyClosed}) survives 
passing to effective action.

\section{BRST formalism}\label{sec:BRST}
We do not have a complete description of symmetry groups $H$ satisfying the properties 
summarized in Section \ref{sec:SummaryOfConditions}. But we do understand the special case, when BV formalism comes from 
BRST formalism.

\subsection{Brief review of BRST formalism}
One starts with the ``classical action'' $S_{\rm cl}$ which is invariant under some gauge symmetry. Let 
$X$ be the ``classical'' space of fields, {\it e.g.} for the Yang-Mills theory the fields are 
$A_{\mu}(x)$. Suppose the gauge symmetry is $H$, with the Lie algebra ${\bf h} = \mbox{Lie}(H)$. We introduce
ghost fields, geometrically:
\begin{align}
\mbox{\tt\small fields and ghosts } & \in {\Pi TH\times X\over H}
\end{align}
where the action of $h_0\in H$ is via right shift on $H$:
\begin{equation}\label{PTHxX}
h_0 (dh,h,x) = (dhh_{0}^{-1}, hh_0^{-1}, h_0.x) 
\end{equation}
We use the coordinates $dh,h$ on $\Pi TH$ 
(denoting ``$dh$'' the coordinate on the fiber $\Pi T_hH$) 
and $x$ on $X$.
Notice that this commutes with $Q_{\rm BRST} = dh {\partial\over\partial h}$. We can always find a representative with 
$h=1$ (i.e. choose $h_0 = h^{-1}$). Then, with the standard notation $c = dh$:
\begin{equation}\label{standard-BRST-operator} 
Q_{\rm BRST} = {1\over 2}f_{AB}{}^Cc^A c^B{\partial\over\partial c^C} + c^A v_A^i{\partial\over\partial x^i} 
\end{equation}
Functions on $\Pi TH\times X\over H$ satisfy:
\begin{align}
    f(dhh_0^{-1}, hh_0^{-1}, h_0 x)\;=\; & f(dh,h,x)
    \\
    (Q_{\rm BRST}f)(dh,1,x)\;=\; & {\partial\over\partial\epsilon}f(dh,\, 1 + \epsilon dh,\, x) = {\partial\over\partial\epsilon}f(dh + \epsilon dh dh,\, 1,\, (1 + \epsilon dh)x)
\end{align}

\subsection{Integration measure}

We assume that $X/H$ comes with some integration measure:
\begin{align}
\mu \;=\; & e^{S_{\rm cl}(x)}
\end{align}
This $\mu$ be understood as an integration measure, i.e. a density of weight $1$, rather 
than a function of $x$. The product of this measure with the canonical measure on $\Pi TH$ 
gives us a measure on $\Pi TH\times X\over H$ which we will call $\mu_{\rm BRST}$. Notice that $Q_{\rm BRST}$ preserves this measure. This can be proven as follows. For any function $f\in\mbox{Fun}\left({\Pi TH\times X\over H}\right)$:
\begin{equation} \int_{\Pi TH\times X\over H}\mu_{\rm BRST}\; Q_{\rm BRST}f \;=\;0 \end{equation}
because $Q_{\rm BRST}$ comes from the canonical odd vector field on $\Pi TH$.

\subsection{Lift of symmetries to BRST configuration space}
Original gauge symmetries can be lifted to the BRST field space as {\em left} shifts on $H$:
\begin{align}
h_0.(dh,h,x) \;=\; & (h_0dh,\, h_0h,\, x)\label{lift-of-symmetry}
\end{align}
These left shifts commute with the right shifts used in Eq. (\ref{PTHxX}), therefore they act
consistently on the factorspace. They also commute with $Q_{\rm BRST}$. 
Moreover, infinitesimal left shifts $L_{\xi}$ are actually BRST-exact:
\begin{align}
\left.{\partial\over\partial t}\right|_{t=0} (e^{t\xi}dh, e^{t\xi}h, x)\;=\;&
\left[(\xi h ){\partial\over\partial dh} \;,\;dh{\partial\over\partial h}\right](dh,h,x)
\\    
\mbox{\tt\small therefore }
L_{\xi} \;=\;&
\left[(\xi h ){\partial\over\partial dh}\;,\;Q_{\rm BRST}\right]
\end{align}
The measure on ${\Pi TH\times X\over H}$ is invariant under left shifts, because it is 
constructed from the canonical measure on $\Pi TH$ which is invariant.

\subsection{BV from BRST}\label{sec:BVfromBRST}
The BV phase space is:
\begin{equation} M = \Pi T^* \left({\Pi TH\times X\over H}\right) \end{equation}
The zero section $\Pi TH\times X\over H$ is a Lagrangian submanifold. It comes with the integration measure, which lifts to a half-density on $M$ of the form $e^{S_{\rm BV}}$, where:
\begin{equation} S_{\rm BV} = S_{\rm cl}(x) + (Q_{\rm BRST}c^A)c^{\star}_A + (Q_{\rm BRST}x^i)x_i^{\star} \end{equation}
where $Q_{\rm BRST}$ is as defined in Eq. (\ref{standard-BRST-operator}).

    We can imagine a more general situation when we have a functional $S_{\rm cl}$ with an odd symmetry $Q_{\rm BRST}$ nilpotent off-shell. But, just to describe the ``standard BRST formalism'', we explicitly break the fields into $c^A$ and $x^i$.

We have realized the gauge algebra $\bf h$ as symmetries of the BRST configuration space as left shifts 
on $\Pi TH$. Since the BV phase space is the odd cotangent bundle, we can further lift them to the 
BV phase space. The symmetry corresponding to the infinitesimal left shift (\ref{lift-of-symmetry}) is generated by the 
BV Hamiltonian:
\begin{align}
H\langle\xi\rangle\;=\; & \Delta(\xi^{\alpha}c^{\star}_{\alpha})
\end{align}
In other words:
\begin{equation}
{\cal F} \mbox{ \tt\small is generated by } c^{\star}_{\alpha}
\end{equation}

\subsection{Form $\Omega$ in BRST formalism}

Although the ``BRST formalism in the proper sense of this word'' requires splitting
$\phi$ into ``physical fields'' $\varphi$ and ``ghosts'' $c$, in many cases such a split
is not required and does not play any role. Let us forget about the split for a moment;
just require that BRST operator is nilpotent off-shell:
\begin{equation}
 Q_{\rm BRST}^2 = 0\;\; \mbox{ \tt\small off-shell}
\end{equation}
But the split into $\varphi$ and $c$ will come back very soon in Section \ref{sec:LiftingGauge}.

\paragraph     {Form $\Omega$ in BRST formalism} is given by the following expression:
\begin{align}
\Omega\;=\;&
\int [d\phi] \exp\left(\; S_{\rm cl} + Q_{\rm BRST}\Psi \quad + \quad d\Psi\;\right)
\label{def-omega-BRST}\\   
d\Omega\;=\; & 0
\end{align}
The family $\Lambda$ is a family of gauge fermions $\Psi(\phi)$. To prove $d\Omega=0$ we use that
$\int [d\phi] Q_{\rm BRST}(\ldots) \;=\; 0$ (no BRST anomaly). Notice that we deform the action
$S_{\rm cl} \;\to \;S_{\rm cl} + Q_{\rm BRST}\Psi$ but do not deform $Q_{\rm BRST}$.

Generally speaking, we would like to treat as ``gauge symmetries'' those $\Psi$ which are 
BRST exact plus equations of motion (here ``equations of motion'' are those derived from
$S_{\rm cl} + Q_{\rm BRST}\Psi$). 
In other words, for any vector field $\zeta$ on the field space and any functional $F$ the following 
$\delta\Psi$ should correspond to a symmetry:
\begin{equation}\label{symmetries} \delta_{F,\zeta}\Psi \quad = \quad Q_{\rm BRST}F \quad +\quad {\cal L}_{\zeta}S_{\rm cl} + [{\cal L}_{\zeta},Q_{\rm BRST}]\Psi \end{equation}
This is just a generic field redefinition plus adding a BRST-exact term\footnote{The second term on the RHS, ${\cal L}_{\zeta}S_{\rm cl} + [{\cal L}_{\zeta},Q_{\rm BRST}]\Psi$, can be interpreted as the term ${\cal L}_{\zeta}(S_{\rm cl} + Q_{\rm BRST}\Psi)$
which vanishes on-shell, plus $Q_{\rm BRST}({\cal L}_{\zeta}\Psi)$ which could in principle be absorbed by a 
shift of $F$.}

\paragraph     {No horizontality}
The form $\Omega$ given by Eq. (\ref{def-omega-BRST}) is not horizontal:
\begin{align}
\iota_{\delta_{\zeta,F}}\Omega\;=\;&
\int [d\phi]\left(
   Q_{\rm BRST}F \;+\; {\cal L}_{\zeta}S_{\rm cl} + 
   [Q_{\rm BRST},{\cal L}_{\zeta}]\Psi
\right) 
\nonumber\\  
& \times \exp\left(S_{\rm cl}(\phi) + 
   Q_{\rm BRST}\Psi + d\Psi
\right) 
\label{iota}
\end{align}
\paragraph     {Cartan form} In order to construct equivariant (and then base) form,
we will need to restrict $\zeta$ and $F$ to belong to some linear subspaces:
\begin{align}
\zeta\;\in\;& {\cal F}_{\rm Vect}\subset \mbox{Vect}(\mbox{\tt\small space of fields }\phi)
\\   
F\;\in\;&{\cal F}_{\rm Fun} \subset \mbox{Fun}(\mbox{\tt\small space of fields }\phi)
\end{align}
We will need to require that these subspaces are such that for any $\zeta\in {\cal F}_{\rm Vect}$ and $F\in {\cal F}_{\rm Fun}$:
\begin{itemize}
\item the following conditions (analogous to  $\{\Phi,\Phi\}=0$) hold:
\begin{align}
{\cal L}_{\zeta}^2 \;=\; & 0
\label{NilpotenceLZeta}
\\    
{\cal L}_{\zeta}F\;=\; & 0
\label{LZetaFIsZero}
\end{align}
\item transformations $\delta_{F,\zeta}$ form a closed Lie superalgebra 
\item the map $\zeta \mapsto [{\cal L}_{\zeta},Q_{\rm BRST}]$ is injective ({\it i.e.} nothing goes to zero)
\end{itemize}
Then the following expression:
\begin{equation}\label{equivariant-omega-in-BRST-formalism} 
\Omega^{\tt C}(\Psi,d\Psi,F,\zeta)\;=\; \int [d\phi] \exp\left(\; 
   S_{\rm cl} \;+\; (Q_{\rm BRST} + d + {\cal L}_{\zeta})\Psi \; + \; F \;
\right) \end{equation}
is annihilated by the Cartan differential.   

\paragraph     {Proof} Eqs. (\ref{NilpotenceLZeta}) and (\ref{LZetaFIsZero}) imply:  
\begin{equation} 
(Q_{\rm BRST} + d + {\cal L}_{\zeta})^2 \;=\; [Q_{\rm BRST}\,,\,{\cal L}_{\zeta}] 
\end{equation}
Therefore:
\begin{align}
& d\;\int [d\phi] \exp\left(
\; S_{\rm cl}(\varphi) \;+\; (Q_{\rm BRST} + d + {\cal L}_{\zeta})\Psi \; + \; F \;
\right)\;=\;
\label{differential-of-equivariant-form}\\   
\;=\;&
\int [d\phi] (Q_{\rm BRST} + d + {\cal L}_{\zeta}) \exp\left(\; 
S_{\rm cl}(\varphi) \;+\; (Q_{\rm BRST} + d + {\cal L}_{\zeta})\Psi \; + \; F \;
\right)\;=\;
\label{differential-equals-to}\\       	
\;=\;&
\int [d\phi] \; \left(
   Q_{\rm BRST} F + {\cal L}_{\zeta}S_{\rm cl} + [Q_{\rm BRST},{\cal L}_{\zeta}]\Psi 
\right) \;\times 
\nonumber\\ 
& \times\exp\left(\; 
   S_{\rm cl}(\varphi) \;+\; (Q_{\rm BRST} + d + {\cal L}_{\zeta})\Psi \; + \; F \;
\right)
\end{align}
In passing from (\ref{differential-of-equivariant-form}) to (\ref{differential-equals-to}) we assumed that both $Q_{\rm BRST}$ and $\zeta$ preserve the measure of
integration $[d\phi]$, in other words $\int [d\phi](Q_{\rm BRST} + {\cal L}_{\zeta})(\ldots)=0$. To complete the proof,
we notice that the last line coincides with Eq. (\ref{iota}).     

\subsection{Lifting the gauge symmetry to BRST formalism}\label{sec:LiftingGauge}
As we have just explained, our $\zeta$ and $F$ are restricted to belong to some subspace
${\cal F}_{\rm Vect}\oplus {\cal F}_{\rm Fun}$, which should satisfy certain conditions. A 
geometrically natural
solution to these conditions can be found in the case of ``traditional'' BRST
formalism where the fields $\phi$ are split into physical fields $\varphi$ and ghosts $c$. It corresponds
to the following choice of $\zeta$ and $F$: 
\begin{align}
F \;=\;& 0
\\    
\zeta^A(\phi) {\partial\over\partial\phi^A} \;=\;& \xi^A{\partial\over\partial c^A}
\end{align}
--- constant (= field-independent) shifts of ghosts. Eq. (\ref{equivariant-omega-in-BRST-formalism}) becomes: 
\begin{equation}
\Omega^{\tt C}(\Psi,d\Psi,\xi)\;=\; \int [d\phi] \exp\left(\; 
   S_{\rm cl} \;+\; Q_{\rm BRST}\Psi + d\Psi +\xi^A{\partial \Psi\over\partial c^A} 
\right) 
\end{equation}

\subsection{Faddeev-Popov integration procedure}\label{sec:Faddeev-Popov}
The naive (or ``standard'') Lagrangian submanifold is:
\begin{align}\label{StandardLagrangianSubmanifold}
\varphi^{\star} = c^{\star} = 0
\end{align}
But the restriction of $S_{\rm BV}$ to this  Lagrangian subnamifold coincides with $S_{\rm cl}$, and therefore 
is degenerate; we cannot integrate. In order to resolve the
degeneracy, we have to deform to another Lagrangian submanifold. However, we face a complication.
It is  desirable to keep the ghost number symmetry. But if we restrict ourselves to only
those Lagrangian submanifolds which are invariant under the ghost number symmetry, then 
the standard one of Eq. (\ref{StandardLagrangianSubmanifold}) is {\em rigid}. It does not
admit deformations. Indeed, the deforming gauge fermion should have ghost number $-1$, 
but there are no fields with negative ghost number. Therefore, there are no deformations.

One solution to this problem is to introduce non-minimal fields (the BRST quartet).
Here we will describe another solution (giving the same answer), based on the consideration 
of families of Lagrangian submanifolds.  
It turns out that exist ghost number 
preserving {\em families} of Lagrangian submanifolds deforming the standard one. We can then
use our form $\Omega$ to integrate over these families, thus obtaining a regularized theory
with ghost number symmetry. Let us outline the construction of such families. 
Let $X$ be the space of fields $\varphi$. Suppose that we have a linear space $V$ and a map
$F\;:\;X\to V$ such that the orbits of the symmetry are transversal to the level set 
$Y = F^{-1}(0)$. This $F$ is called ``gauge fixed condition''. For each $\bar{c}\in \Pi V^*$ we can 
consider the following section of $\Pi T^*X$:
\begin{equation}\label{SectionOfOddCotangentBundle}
s(x) = (F_*(x))^*\bar{c}
\end{equation}
where $F_*(x)\;:\;T_xX\to V$ is the derivative of $F$ at the point $x$. The secion (\ref{SectionOfOddCotangentBundle}) 
defines a Lagrangian submanifold of $\Pi T^* X$. (When $\bar{c}=0$ this is $X\subset \Pi T^*X$.) 
Therefore, we have a family of Lagrangian submanifolds parametrized by elements of $\Pi V^*$. Let 
us integrate our PDF $\Omega$, which in this case is equal to:
\begin{equation}
\Omega(\bar{c},d\bar{c}) \;=\;  \int_{L(\bar{c})} \exp\left(S_{\rm BV} + \langle d\bar{c},F\,\rangle\right)
\end{equation}
over this family. The corresponding integral form (the density on $\Pi V^*$) is:
\begin{equation}
I(\bar{c}) = \int_{L(\bar{c})} \delta(F\,)\,e^{S_{\rm BV}}
\end{equation}
This means that we have to integrate:
\begin{align}
& \int_{\Pi V^*} [d\bar{c}] \int_{L(\bar{c})} \delta(F\,)\,e^{S_{\rm BV}}
\\  
\mbox{\tt\small where } &
\left.S_{\rm BV}\right|_{L(\bar{c})} \;=\; S_{\rm cl}(x) + \langle Q_{\rm BRST} x , (F_*(x))^*\bar{c} \rangle\;=\;
\\    
& \;=\;S_{\rm cl} +  c^A T^i_A\partial_iF^a\bar{c}_a
\end{align}
This is the standard Faddeev-Popov integral (see {\it e.g.} Chapter 8 of \cite{Ramond:1981pw}). The integration
is convergent for an appropriate choice of a contour. 

\paragraph     {Yang-Mills theory} 
\begin{align}
S_{\rm cl} \;=\; & \int d^4x\,\mbox{tr}\, (\partial_{\mu}A_{\nu} - \partial_{\nu}A_{\mu} + [A_{\mu},A_{\nu}])^2
\\
Q_{\rm BRST} A_{\mu} \;=\; & D_{\mu}c
\\
Q_{\rm BRST} c \;=\; & {1\over 2}[c,c]
\end{align}
The Landau gauge corresponds to the following function $F\;:\;X\rightarrow V$, where $V$ is the 
space of functions on the four-dimensional spacetime: 
\begin{equation} F(A) = \partial^{\mu}A_{\mu} \end{equation}
Notice that $V$ (and therefore our family of Lagrangian submanifolds) is now 
infinite-dimensional. In this case $(F_*)^*\bar{c}$ is $A^{\star\mu} = \partial^{\mu}\bar{c}$ and:
\begin{equation} \left.S_{\rm BV}\right|_{L(\bar{c})} \;=\; S_{\rm cl}(x) + D_{\mu}c\,\partial^{\mu}\bar{c}\end{equation}
The integration contour is such that $\bar{c}$ is complex conjugate to $c$. The insertion of $\delta(F\,)$
is usually done by means of a Lagrange multiplier.

\subsection{Conormal bundle to the constraint surface}
Here we will describe Lagrangian submanifolds of the type used in the bosonic string worldsheet 
theory. They provide another solution to the problem discussed in the beginning of Section \ref{sec:Faddeev-Popov}.

\paragraph     {Conormal bundle}
Let $\cal Y$ be some family of submanifolds $Y\subset X$ closed under the action of the gauge symmetry\footnote{this means that if $Y_1\in {\cal Y}$ then for any gauge transformation $h$, $hY_1\in {\cal Y}$}
For each $Y \in {\cal Y}$, the {\em odd conormal bundle} of $Y$ (denoted $\Pi(TY)^{\perp}$) is a subbundle of the odd cotangent bundle $\Pi T^* X|_Y$ which consists of those covectors which evaluate to zero on vectors tangent to $Y$. For each $Y\subset X$, the corresponding odd conormal bundle is a Lagrangian submanifold. Given such a family $\cal Y$, let us define a family of Lagrangian submanifolds in the BV phase space in the following way: for every $Y$, the corresponding Lagrangian submanifold is the odd conormal bundle of $Y$, times the space of $c$-ghosts:
\begin{equation}\label{rotated-lag} 
L(Y) \;=\; \Pi(TY)^{\perp}\times \mbox{[$c$-ghosts]}\;=\; \Pi(TY)^{\perp}\times \Pi {\bf h} 
\end{equation}
\paragraph     {Irreducibility and completeness}
Let us ask the following question: under what conditions the restriction of $S_{\rm BV}$ to each 
$L(Y\,)$ is non-degenerate? Or, in case if it is degenerate, how can we characterize 
the degeneracy? We have:
\begin{equation} 
\left. S_{\rm BV} \right|_{L(Y\,)} \;=\; \left.S_{\rm cl}(\varphi)\;\right|_{\varphi\in Y} \;+\; T_A^ic^A\varphi^{\star}_i\;|_{\varphi^{\star}\in \Pi(T\,Y\,)^{\perp}} 
\end{equation}
The second term $T_A^ic^A\varphi^{\star}_i$ is the evaluation of the covector $\varphi^{\star}$ on the tangent vector $Q\varphi\in TX$.

Let us assume that the restriction of $S_{\rm BV}$ to any $L(Y\,)$ has a critical point, 
and study the quadratic terms in the expansion of $\left. S_{\rm BV} \right|_L$ around that critical point.
To define the perturbation theory, we need already {\em the quadratic terms} to be non-degenerate.
Assuming that the critical point is at $\varphi=0$:
\begin{equation} S_{\rm cl}(\varphi) = k_{ij}\varphi^i\varphi^j + o(\varphi^2) \end{equation}
Suppose that all degeneracies of $S_{\rm cl}$ and of $k$ are due to symmetries (the {\em completeness}). In other words:
\begin{align}
\mbox{ker}\;k\;=\; & \mbox{im}\;\tau
\\
\mbox{\tt\small where } \tau & \;:\; {\bf h} \rightarrow TX
\\
\tau^i(\xi) & \;= \; T^i_A\xi^A
\end{align}
Let $s$ be the quadratic part of $\left. S_{\rm BV} \right|_L$:
\begin{equation}
s\;=\; \left. k_{ij}\varphi^i\varphi^j \right|_{\varphi\in T\,Y} \;+\; \left. \tau^i(c)\varphi^{\star}_i\;\right|_{\varphi^{\star}\in (T\,Y\,)^{\perp}} 
\end{equation}
The degeneracy is characterized by the isotropic subspace of $s$ which we denote $\mbox{ker}\;s$:
\begin{equation} 
\mbox{ker}\;s\;=\; (\mbox{im}\,\tau \cap TY)\oplus \Pi\left((\mbox{im}\,\tau \cap TY)\oplus \mbox{ker}\,\tau\;\;\; \oplus \;\;(\mbox{im}\,\tau\;+TY)^{\perp}\right) 
\end{equation}
Let us make the following assumptions:
\begin{enumerate}
\item The space $\mbox{im}\,\tau \cap T\,Y\,$ is zero, in other words $Y$ is transverse to the orbits of $H$. This is a constraint on the choice of $Y$
\item The next term, $\mbox{ker}\,\tau$, is also zero. This kernel being nonzero corresponds to “reducible” gauge symmetries.
\end{enumerate}
But the last term $\Pi \left(\mbox{im}\,\tau\;+T\,Y\,\right)^{\perp}$ is essentially nonzero. 
It can be identified with the cotangent space to our family:
\begin{equation} 
\left(\mbox{im}\,\tau\;+T\,Y\,\right)^{\perp} \;=\; T^*_Y\left({\cal Y}/H\right) 
\end{equation}
where $\cal Y$ is the moduli space of submanifolds $Y\subset X$. Therefore the quadratic part of $S_{\rm BV}$ is degenerate. However this degeneration is removed by the factor $e^{d\widehat{g}\widehat{g}^{-1}}$. Indeed, in this case:
\begin{equation}
d\widehat{g}\widehat{g}^{-1} \;=\; \varphi^{\star}_i dy^i
\end{equation}
When we integrate over ${\cal Y}/H$, the differentials $dy^i$ span the complement of $\mbox{im}\tau$ in $TX/T\,Y$. Since we require that the family $\cal Y$ be $H$-closed, $\tau$ defines a map ${\bf h} \rightarrow TX/T\,Y$ which we denote $[\tau]$. With these notations:
\begin{equation}\label{factor-tau} 
\left(\mbox{im}\,\tau\;+T\,Y\,\right)^{\perp} = (\mbox{coker}\,[\tau])^* 
\end{equation}
In the case of bosonic string (Section \ref{sec:BosonicStringFamily}) 
$\mbox{dim} \;\mbox{coker}\,[\tau] \;=\; 3g-3$ and $e^{d\widehat{g}\widehat{g}^{-1}}$ contributes 
$\prod_{i=1}^{3g-3} b^{\alpha\beta}\delta g_{\alpha\beta}^{(0)}$.

\section{Integrated vertex operators}\label{sec:IntegratedVertexOperators}
\subsection{Deformations of BV action}
Suppose that we infinitesimally deformed the solution of the Master Equation:
\begin{equation}\label{DeformationOfSBV} S_{\rm BV}(\varepsilon) = S_{\rm BV} + \varepsilon U \end{equation}
In string theory $U$ is called ``integrated vertex operator''.
In order for this to satisfy the Master Equation to the first order in $\varepsilon$ we require:
\begin{equation}\label{MEforU} 
\Delta U\;=\;\rho_{1\over 2}^{-1}\Delta_{\rm can}(\rho_{1\over 2} U)\;=\;
\Delta^{(0)}U + \{S_{\rm BV},U\} = 0
\end{equation}
We will also postulate that $U$ is diffeomorphism invariant, which means in our formalism that:
\begin{equation}\label{UisDiffInv} \{\Delta \Phi,\,U\} =0 \end{equation}
As we deform $S_{\rm BV}$, what happens to $\cal F$?
The answer is simple if $U$ satisfies the Siegel gauge condition:
\begin{equation}\label{Siegel-gauge} \forall \Phi \in {\cal F}\;:\;\{\Phi,U\} = 0 \end{equation}
In this case $\cal F$ remains undeformed.
But if  Eq. (\ref{Siegel-gauge}) is not satisfied, then $\cal F$ should also get deformed.  

\subsection{Deformations of $\cal F$}\label{sec:DeformationOfF}
Eqs. (\ref{MEforU}) and (\ref{UisDiffInv}) imply that $\{U,\Phi\}$ is $\Delta$-closed:
\begin{equation}\label{UFIsClosed}
\Delta\{U,\Phi\}=0
\end{equation}
Let us also require that it is $\Delta$-exact:
\begin{align}
\exists\; a_U\;:\;{\cal F}\to \mbox{Fun}(M) & : 
\\    
\{U,\Phi\} & \;=\;- \Delta a_U(\Phi)  \label{DefAPhi}
\\    
\mbox{\tt\small and } [\Delta\Phi\langle \xi\rangle\,,\,a_U(\Phi\langle\eta\rangle)] & \;=\;a_U(\Phi\langle[\xi,\eta]\rangle)   \mbox{ \tt\small ($\bf h$-invariance of $a_U$)}
\label{ActionOfDiffeomorphismsOnFIsUndeformed}
\end{align}
Under these assumptions we can deform: 
\begin{align}
\Phi\;\mapsto\; & \Phi + \varepsilon a_U(\Phi)\label{DiffeomorphismsAraUndeformed}
\\   
\mbox{\tt\small so that } & (\Delta + \varepsilon\{U\,,\,\_\})(\Phi + \varepsilon a_U(\Phi)) = \Delta \Phi
\label{BVHamiltonianOfDiffStaysUndeformed}
\end{align}
In other words:
\begin{itemize}
\item the space $\cal F$ does deform, according to Eq. (\ref{DiffeomorphismsAraUndeformed}), but the action of diffeomorphisms 
remains the same
\end{itemize}
In particular, the BV Hamiltonian of diffeomorphisms $\Delta \Phi$ stays undeformed (Eq. (\ref{BVHamiltonianOfDiffStaysUndeformed}))  

Is it true that the deformed $\Phi$ remain in involution modulo $\Delta$-exact? 
Notice that the deformed $\{\Phi,\Phi\}$ is automatically $\Delta + \varepsilon\{U\,,\,\_\}$-closed under already taken assumptions:
\begin{equation}\label{DefDeltaDefPhiPhi}
(\Delta + \varepsilon\{U\,,\,\_\})\{\Phi + \varepsilon a_U(\Phi)\,,\,\Phi + \varepsilon a_U(\Phi)\} \;=\; 2\{\Delta\Phi\,,\, \Phi + \varepsilon a_U(\Phi)\} = 0 
\end{equation}
Opening the parentheses, we derive that the expression 
$2 \{\Phi\,,\,a_U(\Phi)\} - \{U\,,\,q(\Phi)\}$ is $\Delta$-closed:
\begin{equation} 
\Delta\Big(2 \{\Phi\,,\,a_U(\Phi)\} - \{U\,,\,q(\Phi)\}\Big) = 0 
\end{equation}
Let us assume that it is also $\Delta$-exact:
\begin{equation} 
2 \{\Phi\,,\,a_U(\Phi)\} - \{U\,,\,q(\Phi)\}\;=\;\Delta q'(\Phi) 
\end{equation}
(the validity of this assumption depends on the cohomology of $\Delta$). 
Therefore, to the first order in the bosonic infinitesimal parameter $\varepsilon$:
\begin{equation}
\{\Phi + \varepsilon a(\Phi) \;,\; \Phi + \varepsilon a(\Phi) \} = (\Delta + \varepsilon\{U,\_\})(q(\Phi) + \varepsilon q'(\Phi))
\end{equation}
\commentstarts{\small
This equation can also be derived immediately from Eq. (\ref{DefDeltaDefPhiPhi}) under the assumption that the
ghost number 3 cohomology of $\Delta + \varepsilon\{U\,,\,\_\}$ vanishes
}\commentends

\noindent
Therefore the condition of being in involution persists, but with deformed $q$: 
\begin{equation} 
q \mapsto q + \varepsilon q' 
\end{equation}
Remember that the construction of equivariant form $\Omega^{\tt C}$ requires solving the equation:
\begin{equation}
\Delta_{\rho_{1/2}} a(\xi) + {1\over 2} \{a(\xi),a(\xi)\} \;=\;\xi \quad (\;=\;\Delta\Phi\langle\xi\rangle\;)
\end{equation}
The solutions deforms:	
\begin{equation}
\begin{array}{cccc}
a(\xi) \quad\mapsto\quad a(\xi)\; & + \quad \varepsilon a'_U\langle\xi\rangle \; & - \quad \varepsilon q'_U(\xi) \;& + \ldots
\cr
 & \mbox{\tt\small linear in } \xi \qquad & \mbox{\tt\small quadratic in } \xi &  
\end{array}
\end{equation}
And the string measure deforms: 
\begin{equation}
\begin{array}{ccccc}
\Omega^{\tt C}(L,\Psi,\xi)\;=\; \int_L \exp\Big( & S_{\rm BV} & \;+\;\Psi\;+\; & a(\xi) & \Big)
\cr
 & \mbox{\tt\small becomes} & & \mbox{\tt\small becomes} &
\cr
 & S_{\rm BV} + \varepsilon U & & a +\varepsilon a_U' -\varepsilon q_U' + \ldots &
\end{array}
\end{equation}
In the base form $\Omega^{\tt base}$ we substitute for $\xi$ the curvature of the connection --- the 2-form. 
In other words, $a(\xi)$ becomes a {\em two-form} on $\rm LAG$.
Therefore a vertex operator is actually {\em an inhomogeneous PDF} on $\rm LAG$ (and not just a
function).
If the theory has ghost number, then $U$ has ghost number zero and $a(\xi)$ has ghost number $-2$;
the sum of the ghost number and the degree of the form is zero.
Lower ghost number components of vertex operators were recently used in the context 
of pure spinor formalism in \cite{Berkovits:2016xnb}.

\subsection{Exact vertex operators}
Let us consider the case when $U$ is $\Delta$-exact:
\begin{equation}
U = \Delta W
\end{equation}
Let us also assume that $W$ is $H$-invariant: 
\begin{equation}
\{\Delta \Phi, W\} \;=\;0
\end{equation}
In this case, we can take:
\begin{equation}
a_U(\Phi) \;=\; - \{W,\Phi\}
\end{equation}
We observe that the resulting deformation of the equivariant form is the Lie derivative along $\{W,\_\}$:
\begin{align}
{\cal L}_{-\{W,\_\}} \Omega^{\tt C} = \Omega^{\tt C}\langle \Delta W - \{W,\Phi\} \rangle
= \Omega^{\tt C}\langle U + a_U(\Phi)\rangle
\end{align}
This means that: 
\begin{itemize}
\item deforming with an exact integrated vertex is equivalent to an infinitesimal change of the integration
cycle
\end{itemize}
\commentstarts{\small
To construct the base form from the Cartan form, we need to choose a connection. When deforming with an
exact vertex, we need to also adjust the connection by carrying it along with $\{W,\_\}$
}\commentends

\subsection{Relaxing Siegel gauge}
Let us deform $S_{\rm BV}$ by adding an integrated vertex operator not in the Siegel gauge.
In this case we have to use the full base form $\Omega^{\tt base}$ from Section \ref{sec:BaseForm}.
Suppose that there exists $W$ such that $U + \Delta W$ {\em is} in Siegel gauge:
\begin{equation}
\{\Phi\,, \, U + \Delta W\} = 0
\end{equation}
In practice $W$ could be complicated. But if it exists, then this implies that we can satisfy
the Siegel gauge by slightly deforming the integration cycle. 

Usually in string theory models, having the Siegel gauge satisfied on a family of Lagrangian 
submanifolds implies that $\Phi$ vanishes on Lagrangian submanifolds in this family; then our form $\Omega^{\rm base}$
reduces to the standard string theory measure. (In the case of bosonic string this is discussed in
Section \ref{sec:BosonicStringIntegration}.) Combined with the observation that the Siegel gauge can be satisfied by deforming the 
contour, this implies that the amplitudes obtained by integrating our $\Omega^{\tt base}$ give the same result as 
the standard prescription for string amplitudes.

\section{Bosonic string}
Now we will explain how this formalism can be applied to the bosonic string worldsheet theory.

\subsection{Solution of Master Equation}
The fundamental fields of the worldsheet theory are: matter fields $x^m$, complex structure $I^{\alpha}_{\beta}$ and the 
diffeomorphism ghosts $c^{\alpha}$. The matter part of the action depends on matter fields and complex structure:
\begin{equation}
S_{\rm mat}[I,x] = {1\over 2}\int \partial_+ x^m \wedge \partial_- x^m
\end{equation}
Following the general scheme, we find the solution of the Master Equation:
\begin{equation}\label{BosonicMasterAction}
S_{\rm BV} = S_{\rm mat} \;+\; \int \; \langle {\cal L}_c I , I^{\star}\rangle + \langle {\cal L}_c x , x^{\star}\rangle + {1\over 2}\langle [c,c],c^{\star}\rangle
\end{equation}
There is a constraint: $I^2 = -1$, or in components: $I^{\alpha}_{\beta}I^{\beta}_{\gamma} = - \delta^{\alpha}_{\gamma}$. The antifield $I^{\star}$ can be 
identified with a symmetric tensor $I^{\star}_{\alpha\beta} = I^{\star}_{\beta\alpha}$ with only nonzero components $I^{\star}_{++}$ and $I^{\star}_{--}$ (which are also 
denoted $I^{\star}_{zz}$ and $I^{\star}_{\bar{z}\bar{z}}$.) The coupling to $\delta I$ is defined as follows:
\begin{equation}\label{PairingI}
\int \langle I^{\star}, \delta I\,\rangle \;=\; \int \, dz^{\alpha}\wedge dz^{\beta}\;I^{\star}_{\alpha\gamma}\delta I^{\gamma}_{\beta}
\end{equation}

\subsection{Family of Lagrangian submanifolds}\label{sec:BosonicStringFamily}
\paragraph     {Motivation for changing polarization}
Notice that the dependence of $S_{\rm BV}$ on the antifields (letters with $\star$) is at most linear. Indeed, this $S_{\rm BV}$ corresponds to ``just the usual BRST operator'' of the form $c^At_A + {1\over 2}f^A_{BC}c^Bc^C{\partial\over\partial c^A}$.

It would seem to be natural to choose the Lagrangian submanifold setting all the antifields to zero. However, the restriction of $S_{\rm BV}$ to this Lagrangian submanifold (i.e. $S_{\rm cl}$) turns out to be problematic from the point of view of quantization (the Nambu-Goto string). The standard approach in bosonic string is to switch to a different Lagrangian submanifold so that the restriction of $S_{\rm BV}$ to this new Lagrangian submanifold is quadratic. However there is some price to pay: BRST operator is only nilpotent on-shell.

\paragraph     {New polarization}
Let us choose some reference complex structure $I^{(0)}$ and parametrize the nearby complex 
structures by their corresponding Dolbeault cocycles, which we denote $b^{\star}$. Locally it is 
possible to choose $I = \left(\begin{array}{cc} i & 0 \cr 0 & -i \end{array}\right)$. To summarize:
\begin{align}
I \;=\; & I(b^{\star})
\\
b^{\star}\;\in\; & H^1_{\overline{\partial}}(T^{1,0})
\\
I^z_{\bar{z}}(b^{\star}) \;=\; & (b^{\star})^z_{\bar{z}} + o(b^{\star})
\\
I^z_z(b^{\star})\;=\; & i + o(b^{\star})
\end{align}
The other components are:
\begin{equation}
I^{\bar{z}}_{\bar{z}} = \overline{I^z_z}\;,\quad I^{\bar{z}}_z = \overline{I^z_{\bar{z}}}
\end{equation}
And we rename $I^{\star}$ as $b$:
\begin{equation}
b_{\alpha\beta} \;=\; I^{\star}_{\alpha\beta}
\end{equation}
We have just changed the polarization; $I^{\star}$ is now a field (called $b$) and $I$ an antifield (called $b^{\star}$). The action can be written in the new coordinates:
\begin{align}
S_{\rm BV}\;=\; & S_{\rm mat}[I(b^{\star})\,,\, x] \;+\; \int\;\langle {\cal L}_c(I(b^{\star})), \, b\rangle + \langle {\cal L}_c x,\, x^{\star} \rangle + \left\langle {1\over 2}[c,c],\, c^{\star}\right\rangle\label{BVBosonic}
\end{align}

\paragraph     {Family of Lagrangian submanifolds}
We now choose the Lagrangian submanifold in the following way:
\begin{align}
b^{\star} = x^{\star} = c^{\star} = 0\label{LagrangianAlongB}
\end{align}
On this Lagrangian submanifold the action is quadratic. In particular:
\begin{equation}
\int \langle b,{\cal L}_c(I(b^{\star}))\rangle\quad\mbox{\tt\small becomes}\quad \int (b_{++}\partial_-c^+ + c.c)
\end{equation}

\paragraph     {BRST structure}
The BRST operator $Q_{\rm BRST}$ of the bosonic string can be understood as follows. We expand $S_{\rm BV}$ in powers of the antifields $x^{\star}, c^{\star}, b^{\star}$ and consider only the linear term. The corresponding Hamiltonian vector field preserves the Lagrangian submanifold and is the symmetry of the restriction of $S_{\rm BV}$ on the Lagrangian submanifold. There are also higher order terms, because the dependence on $b^{\star}$ is nonlinear; $\int b^{\star\,\gamma}_{\alpha}T_{\gamma\beta}\;dz^{\alpha}\wedge dz^{\beta}$ is just the linear approximation. In constructing the $Q_{\rm BRST}$ we simply neglect those higher order terms. This leads to $Q_{\rm BRST}$ being nilpotent only on-shell. The explicit formula for $Q_{\rm BRST}$ can be read from Eq. (\ref{BVBosonic}):
\begin{align}
Q_{\rm BRST} \;=\; &   T_{\alpha\beta}{\partial\over\partial b_{\alpha\beta}} + (c^{\alpha}\partial_{\alpha}x^m) {\partial\over\partial x^m} + {1\over 2}[c,c]^{\alpha}{\partial\over\partial c^{\alpha}}
\end{align}

\paragraph     {Ghost numbers}
\begin{tabular}{| c | c | c | c | c | c |}
\hline
$x$ & $x^{\star}$ & $I$ & $b$ & $c$ & $c^{\star}$
\\   \hline
$0$ & $-1$ & $0$ & $-1$ & $1$ & $-2$
\\   \hline
\end{tabular}

\subsection{Integration over the family of Lagrangian submanifolds}\label{sec:BosonicStringIntegration}
Here we will first use our prescription to construct the equivariant analogue of $\Omega$, depending 
on some motivated choice of the subspace ${\cal F}\subset \mbox{Fun}(M)$. We will then implement the standard 
procedure to construct a closed form on $H\backslash {\rm LAG}$.

\paragraph     {Choice of $\cal F$}
We will here make use of the standard choice of $\cal F$ always applicable in the BRST case as we explained
in Section \ref{sec:BVfromBRST}. We will choose $\cal F$ so that $\Pi\Delta {\cal F}$ is the algebra of diffeomorphisms of the worldsheet:

\begin{equation}\label{DefPhi} \Phi = \int c^*_{\alpha} \xi^{\alpha} \end{equation}
where $\xi^{\alpha} = \xi^{\alpha}(z,\bar{z})$ is a vector field on the worldsheet.

\paragraph     {Equivariant form $\Omega^{\cal C}_{{\rm Vect}(\Sigma)\oplus{\rm Weyl}}$}
The resulting equivariant form is:
\begin{equation} \Omega^{\tt C}_{{\rm Vect}(\Sigma)\oplus {\rm Weyl}}({\bf t}) = \int_{gL} \exp\left( S_{\rm BV} + \int \langle dI\,,\, b\rangle + \int c^{\star}_{\alpha} {\bf t}^{\alpha} \right) \end{equation}
where the term $\int \langle dI\,,\, b \rangle$ comes from $dg g^{-1}$ and $\int c^{\star}_{\alpha}{\bf t}^{\alpha}$ from $\Phi$ of Eq. (\ref{DefPhi}). The pairing 
$\langle dI\,,\,b\rangle$ is as in Eq. (\ref{PairingI}).

\paragraph     {Recovery of the standard approach}
Usually in the literature, the integration cycle is chosen so that $c^{\star}=0$, and therefore 
the term $\int c^{\star}_{\alpha}{\bf t}^{\alpha}$ vanishes. Moreover, we do not even need to do the horizontal projection of 
$dgg^{-1}$. This is a consequence of the following general statement. Suppose that $\{\Phi,\Phi\}=0$ 
and the integration cycle in the moduli space of Lagrangian submanifolds is such that:
\begin{align}
\Phi|_{gL}=\; & 0\label{PhiIsZero}
\end{align}
In this case the construction simplifies:
\begin{equation} \Omega_{{\rm Vect}(\Sigma)} = \Omega \end{equation}
We can now use the Baranov-Schwarz transform (Section \ref{sec:BaranovSchwarz}) and interpret the integration 
of $\Omega$ as the integration of $e^S$ over some new Lagrangian submanifold, which can be described 
as follows. Consider {\em any} $3g-3$-dimensional surface in the space of metrics $g_{\cdot\cdot}$ 
parametrized by $(s_1,\ldots,s_{3g-3})$:
\begin{equation} S\subset {\rm MET}\;,\quad \mbox{dim}S = 3g-3 \end{equation}
(no need to require any holomorphicity). Let us consider a submanifold
\begin{equation} {\cal N}_S\subset \Pi T^*{\rm MET} \end{equation}
defined as follows: it is the bundle over $S$ whose fiber at a point $s\in S$ consists of the 
subspace of $\Pi T^*_s{\rm MET}$ orthogonal to the tangent space to $S$ at that point. Notice that 
this submanifold is Lagrangian. We will promote ${\cal N}_S$ to a Lagrangian submanifold $\widehat{\cal N}_S$ in the 
BV phase space of the bosonic string by adding $x$ and $c^{\cdot}$ (and keeping $x^{\star}=0$ and $c^{\star}_{\cdot}=0$). 
We have:
\begin{equation}\label{BosonicStringSingleLagrangian} \int_S\Omega = \int_{\widehat{\cal N}_S} e^{S_{\rm BV}} \end{equation}
Notice that we can use $\Omega$ instead of $\Omega_{\cal B}$ because Eq. (\ref{PhiIsZero}) is satisfied in this case. 
Indeed, $\Phi\langle\xi\rangle = \int \xi^{\alpha}c^{\star}_{\alpha}$ and we choose the Lagrangian submanifolds so that $c_{\cdot}^*=0$; this 
proves Eq. (\ref{PhiIsZero}).

When some part of $\widehat{\cal N}_S$ contains a gauge-trivial direction $\dot{g}_{\alpha\beta} = 2\nabla_{(\alpha}\xi_{\beta)}$, then the integral of $e^{S_{\rm BV}}$ over that part is automatically zero. Indeed, in this case all the $b^{\cdot\cdot}\in\Pi T^*_s{\rm MET}$ orthogonal to the tangent space to $S$ satisfy in particular $b^{\alpha\beta}\nabla_{\alpha}\xi_{\beta}=0$ and therefore the integral of $e^{S_{\rm BV}}$ over $c^{\cdot}$ will give zero (because of the zero mode $c^{\alpha}\simeq \xi^{\alpha}$). In this sense, $e^{S_{\rm BV}}$ is a ``base integral form''.

\paragraph     {Standard integration cycle}
We will now discuss the ``usual'' (in the bosonic string theory) integration cycle on the moduli space of 
Lagrangian submanifolds. Our choice of $\cal F$ is such that ${\bf h}=\Pi\Delta{\cal F}$ is the algebra of diffeomorphisms 
of the worldsheet. This allows us to construct the base form on the space of Lagrangian submanifolds 
modulo diffeomorphisms.

On a Lagrangian submanifold from the standard family the metric $g_{\alpha\beta}$ (same thing as $b^{\star\alpha\beta}$) is fixed, 
and the path integral goes over $b^{\alpha\beta},c^{\alpha},x^m$. This picture explains why factorization by the 
action of $H$ results in closed integration cycles:\\
\includegraphics[scale=0.75]{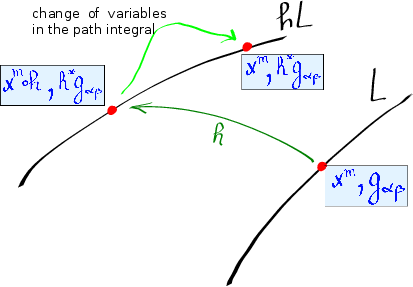}\\
Here $h$ is a large diffeomorphism; the integration cycle is a $3g-3$-dimensional
family of metrics $g_{\alpha\beta}$; it only becomes closed after we make an identification
of $g_{\alpha\beta}$ and $h^*g_{\alpha\beta}$.

\section{Topologically twisted $N=2$ model}\label{sec:N2}
In this Section we will give an example of the contructrion of $\cal F$ in the case which is
not covered by the scheme described in Section \ref{sec:BVfromBRST}. 

\subsection{BV action}
Consider the topologically B-twisted $N=2$ superconformal theory \cite{Klemm:Intro}.
Let us restrict to the case of flat target space. The fields are: 

\begin{tabular}{|r|l|}
\hline
bosons: & complex scalars $x^a$ and $\overline{x^a}$ 
\\ \hline 
fermions: & one-form $\rho^a$  and scalars $\vartheta_a$ and $\overline{\eta^a}$
\\ \hline
\end{tabular}

\vspace{10pt}
\noindent
The ``classical'' action is:
\begin{equation}\label{ActionBModel}
S_{\rm cl} \;=\; \int_{\Sigma}-{1\over 2} \, d\overline{x^a}\wedge *d x^a -\rho^a\wedge (*d\overline{\eta^a} + d\vartheta_a)
\end{equation}
The BV action is:
\begin{equation}\label{BVActionBModel}
S_{\rm BV} \;=\; S_{\rm cl} \,+\,\int_{\Sigma}\; \overline{\eta^a}\overline{x}_{\overline{a}}^{\star} \,-\,{1\over 2} \langle dx^a,\rho^{\star}_a\rangle
\end{equation}

\subsection{Action of diffeomorphisms}
Let us consider the infinitesimal diffeomorphism (= vector field) of the worldsheet:
\begin{equation}
v^{\scriptsize\rm L}(z^{\scriptsize\rm L},z^{\scriptsize\rm R}){\partial\over\partial z^{\scriptsize\rm L}} + v^{\scriptsize\rm R}(z^{\scriptsize\rm L},z^{\scriptsize\rm R}){\partial\over\partial z^{\scriptsize\rm R}}
\end{equation}
The corresponding BV generation function is ${\cal V}\langle v\rangle\;=\;\{S_{\rm BV}\,,\,\Phi\langle v\rangle\}$ where:
\begin{align}
\Phi\langle v\rangle \;=\; & \int_{\Sigma} ({\cal L}_v\overline{x^a})\overline{\eta}^{\star}_{\bar{a}} - ({\cal L}_{(v_L - v_R)}\overline{x^a}) \vartheta^{\star a} - 2 (\iota_v\rho^a)x^{\star}_a + \iota_{\rho^{\star}_a}\iota_v\vartheta^{\star a}\label{Phi}
\end{align}
\commentstarts{\small
The last term $\iota_{\rho^{\star}_a}\iota_v\vartheta^{\star a}$ requires explanation. As $\vartheta$ is a scalar field, its antifield $\vartheta^{\star}$ is a 2-form. The expression $\iota_v\vartheta^{\star}$ is the usual contraction of a vector field $v$ with the 2-form $\vartheta^{\star}$. Finally, $\iota_{\rho^{\star}_a}\iota_v\vartheta^{\star}$ should be understood in the following way. We interpret $\rho^{\star}$ as a vector-valued two-form, i.e. a section of $(T\otimes \Omega^2)\Sigma$. We contract its vector index with the single remaining covector index of $\iota_v\vartheta^{\star}$. What remains is a two-form, which is just integrated over $\Sigma$.}
\commentends

\noindent Explicitly:
\begin{align}
{\cal V}\langle v\rangle \;=\; & \{S_{\rm BV}\,,\,\Phi\langle v\rangle\} \;=\; 
\nonumber \\  
\;=\; &
\int_{\Sigma} ({\cal L}_v x^a)x^{\star}_a + ({\cal L}_v \overline{x^a})\overline{x}^{\star}_{\overline{a}} + \langle({\cal L}_v\rho^a),\rho^{\star}_a\rangle + ({\cal L}_v\overline{\eta^a})\overline{\eta}^{\star}_{\overline{a}} + ({\cal L}_v\vartheta_a)\vartheta^{a\star} +
\nonumber \\
& \;\;\; + \; d\overline{x^a}\wedge *d(\iota_v\rho^a) - ({\cal L}_v\overline{x^a})d*\rho^a - ({\cal L}_{(v_L-v_R)}\overline{x^a})d\rho^a
\label{GeneratorOfDiffeomorphisms}
\\
\;=\; & \int_{\Sigma} ({\cal L}_v x^a)x^{\star}_a + ({\cal L}_v \overline{x^a})\overline{x}^{\star}_{\overline{a}} + \langle({\cal L}_v\rho^a),\rho^{\star}_a\rangle + ({\cal L}_v\overline{\eta^a})\overline{\eta}^{\star}_{\overline{a}} + ({\cal L}_v\vartheta_a)\vartheta^{a\star} -
\nonumber \\
& \;\;\; - *d\overline{x^a} \wedge {\cal L}_v\rho^a - ({\cal L}_v \overline{x^a}) d*\rho^a
\nonumber
\end{align}
Identities useful in proving this:
\begin{align}
& \left\{\; \int_{\Sigma}\langle dx^a,\rho^{\star}_a\rangle \;,\; \int_{\Sigma}(\iota_v\rho^b)x_b^{\star}\;\right\}\;=\; 
\nonumber\\   
\;=\; & \int_{\Sigma}({\cal L}_v x^a) x_a^{\star}\;+\; \langle (d\iota_v \rho^a),\rho_a^{\star} \rangle
\\[8pt]
& \left\{\; \int_{\Sigma} \overline{\eta^a}\overline{x}^{\star}_{\bar{a}} \;,\; \int_{\Sigma} ({\cal L}_v\overline{x^a})\overline{\eta}^{\star}_{\overline{a}} - ({\cal L}_{(v_L - v_R)} \overline{x^a})g_{\bar{a}b}\vartheta^{\star b} \;\right\}\;=\; 
\nonumber\\   
\;=\;& \int_{\Sigma}\;({\cal L}_v\overline{\eta^a})\overline{\eta}^{\star}_{\bar{a}} \,+\,({\cal L}_v\overline{x^a})\overline{x}^{\star}_{\bar{a}} \,-\,({\cal L}_{(v_L - v_R)}\overline{\eta^a})g_{\bar{a}b}\vartheta^{\star b}
\\[8pt]
& \left\{\; S_{\rm cl} \;,\; \int_{\Sigma}\;\iota_{\rho^{\star}_a}\iota_v\vartheta^{\star a} \;\right\} \;=\; 
\nonumber\\   
\;=\;& \left\{\; \int_{\Sigma}\; - *d\overline{\eta^a}\wedge \rho^a + \vartheta_a d\rho^a \;,\; \int_{\Sigma}\;\iota_{\rho^{\star}_a}\iota_v\vartheta^{\star a} \;\right\}\;=\;
\nonumber\\  
\;=\; & \int_{\Sigma}\;\langle (\iota_v d\rho^a) , \rho^{\star}_a\rangle\;+\;(d\vartheta_a + * d\overline{\eta^a}) \wedge \iota_v\vartheta^{\star a} \;=\;
\nonumber\\    
\;=\; & \int_{\Sigma}\;\langle (\iota_v d\rho^a) , \rho^{\star}_a\rangle \;+\; ({\cal L}_v\vartheta_a)\vartheta^{\star a} \;+\; ({\cal L}_{(v_L-v_R)}\overline{\eta^a})\vartheta^{\star a}
\end{align}

\subsection{Closedness of ${\cal F}$ under $[\_,\_]$}
We need to prove:
\begin{equation}
\{\{S_{\rm BV},\Phi\langle w\rangle\},\Phi\langle v\rangle\} = \Phi\langle [w,v]\rangle
\end{equation}
The only nontrivial computation is the bracket of the last term in Eq. (\ref{Phi}) with the last line of (\ref{GeneratorOfDiffeomorphisms}):
\begin{align}
& \left\{ \int_{\Sigma} \iota_{\rho^{\star}_a}\iota_v\vartheta^{\star a}\,,\, \int_{\Sigma} - *d\overline{x^a} \wedge {\cal L}_w\rho^a - ({\cal L}_w \overline{x^a}) d*\rho^a \right\} \;=\;
\\
\;=\; & \int_{\Sigma} - * d\overline{x^a}\wedge {\cal L}_w\iota_v\vartheta^{\star a} \;-\; ({\cal L}_w\overline{x^a})d * \iota_v\vartheta^{\star a} \;=\;
\\
\;=\; & \int_{\Sigma} - * d\overline{x^a}\wedge {\cal L}_w\iota_v\vartheta^{\star a} \;-\; *d({\cal L}_w\overline{x^a}) \iota_v\vartheta^{\star a}\label{CommutatorA}
\end{align}
and the bracket of the term $- ({\cal L}_{(v_L - v_R)}\overline{x^a}) \vartheta^{\star a}$ in Eq. (\ref{Phi}) with $\widehat{\cal L}_w$:
\begin{align}
& \left\{ \int_{\Sigma} -\iota_v *d\overline{x^a} \vartheta^{\star a}\;,\;\widehat{\cal L}_w \right\}\;=\;
\\
\;=\; & \left\{ \int_{\Sigma} -*d\overline{x^a}\wedge\iota_v \vartheta^{\star a}\;,\;\widehat{\cal L}_w \right\}\;=\;
\\
\;=\; & \int_{\Sigma} * d\overline{x^a}\wedge \iota_v{\cal L}_w\vartheta^{\star a} \;+\; *d({\cal L}_w\overline{x^a}) \iota_v\vartheta^{\star a}\label{CommutatorB}
\end{align}
Eqs. (\ref{CommutatorA}) and (\ref{CommutatorB}) combine into $\int_{\Sigma} *d\overline{x^a}\wedge \iota_{[w,v]}\vartheta^{\star a}$ which is the required term 
$- ({\cal L}_{([w,v]_L - [w,v]_R)}\overline{x^a})\vartheta^{\star a}$ in $\Phi\langle [w,v]\rangle$

\section{Worldsheet with boundary}\label{sec:WorldsheetWithBoundary}
The consideration of this Section is not rigorous for the reason explained in Section \ref{sec:ReviewBV}. 
We hope that more rigorous treatment could be obtained by an axiomatic approach along the 
lines of \cite{Cattaneo:2015vsa}. 

Consider string worldsheet theory on a flat disk $D$. Insertion of some operators inside the 
disk creates a state on the boundary:\\
\includegraphics[scale=0.5]{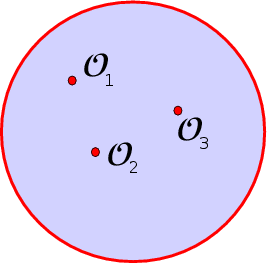}\\
Let us consider a Riemann surface with a boundary, and insert our flat disk:\\
\includegraphics[scale=0.3]{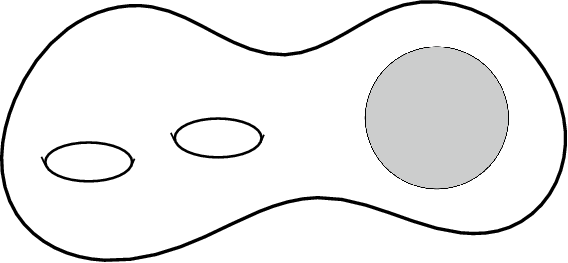}
\hspace{10pt}
\includegraphics[scale=0.3]{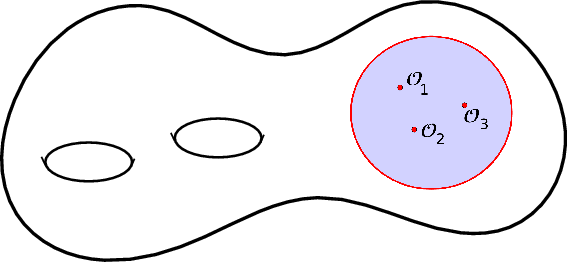}\\
Consider the variations of the Lagrangian submanifold (for example, metric), 
limited inside  some  compact region on the Riemann surface:\\
\includegraphics[scale=0.5]{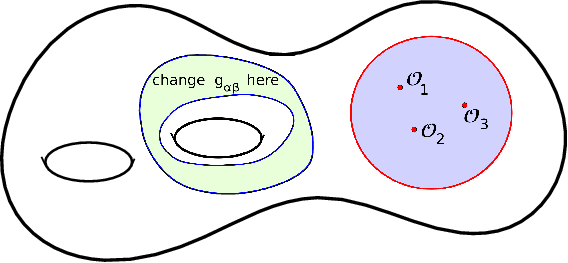}\\
As we vary the metric, the state on the boundary (marked red) remains the same state in the 
same theory. This is possible because we do not change the theory in the region of insertion.

If we limit the variations of the Lagrangian submanifold in this way, then we can construct the 
form $\Omega$ (and its base analogue) in the same way as we did on the closed Riemann surface. In 
this case the form $\Omega$ is not closed:
\begin{equation}\label{DOmegaPsi} 
d\Omega^{\tt base}_{\Psi} = \Omega^{\tt base}_{Q_{\rm BRST}\Psi} 
\end{equation}
We will derive this formula in a moment.
If we restrict ourselves with the insertions of only physical operators $\Psi$, then the form 
$\Omega^{\tt base}_{\Psi}$ {\em is} closed.

\subsection{Variation of the wave function}
Let us study the quantum theory on a region $\Sigma\backslash D$ (the complement of the disk on the Riemann 
surface); its boundary is $-\partial D$. On the picture $D$ is painted blue. Notice that there are 
no operator insertions inside $D$.

Let us bring our theory to some first order formalism, so that the Lagrangian is of the form 
$p\dot{q} - H$. Fix the Lagrangian submanifold $L$. Fix some polarization in the restriction of the 
fields to $\partial D$, for example the standard polarization where the leaves have constant $q$. Let 
us consider the wave function:
\begin{align}
\Psi(q) \;=\; & \int_L \rho_{1/2}\label{DefPsi}
\end{align}
— the integration is over the field configurations inside $D$, and $q$ enters through boundary 
conditions. More generally, let $\Psi_{\cal O}(q)$ denote the wave function obtained by the path 
integration with the insertion of some operator $\cal O$ with compact support, not touching the 
boundary:
\begin{align}
\Psi_{\cal O}(q) \;=\; & \int_L {\cal O}\rho_{1/2}\label{DefPsiWithInsertion}
\end{align}
Any functional $\Xi$ on the odd phase space, with compact support on the worldsheet, determines 
an operator insertion, just by its restriction on the Lagrangian submanifold. 
We will denote it $\Xi_0$:
\begin{equation}
\Xi_0 =\Xi|_L
\end{equation}

\refstepcounter{Theorems}
\paragraph     {Theorem \arabic{Theorems}:\label{theorem:VariationOfPsiGeneral}} 
Suppose that $\rho_{1\over 2}$ satisfies the Master Equation. The expansion of $S_{\rm BV}$ around the 
Lagrangian submanifold in powers of antifiedls defines some $Q_{\rm BRST}$. Then:
\begin{equation}\label{VariationOfPsiGeneral}
\Psi_{Q_{\rm BRST} \Xi_0}(q) \;=\; - \int_L {\cal L}_{\{\Xi,\_\}} \rho_{1\over 2}
\end{equation}
where $\Xi_0$ is the restriction of $\Xi$ to $L$ and ${\cal L}_{\Xi}$ is the Lie derivative of 
the half-density $\rho_{1\over 2}$ along the Hamiltonian vector field $\{\Xi,\_\}$ (see Eq. (\ref{ViaLieDerivative})). Notice: 
\begin{itemize}
\item  Eq. (\ref{VariationOfPsiGeneral}) is not affected by the presence of a boundary
\end{itemize}

\paragraph     {Proof:} Let us introduce some Darboux coordinates near $L$, so that $L$ is given 
by the equation $\phi^{\star} = 0$. Let us expand $\Xi$ in powers of $\phi^{\star}$. Let us first 
assume that only a constant in antifields term is present:
\begin{equation}
\Xi = \Xi_0(\phi)
\end{equation}
In this case the Lie derivative is equivalent to inserting $Q_{\rm BRST}\Xi_0$ into the path 
integral, giving Eq. (\ref{VariationOfPsiGeneral}).

If $\Xi$ also depends on antifields, then we have to be careful restricting ourselves to such $\Xi$ 
that the Lie derivative ${\cal L}_{\Xi} \rho_{1\over 2}$ is well-defined, because otherwise Eq. (\ref{VariationOfPsiGeneral}) does not make 
sense. This assumption must include the vanishing of the integration by parts:
\begin{equation}
\int_L {\partial\over\partial\phi}{\partial\over\partial\phi^{\star}} \left(\Xi_1\phi^{\star} \rho_{1\over 2}\right) = 0
\end{equation}
This is equivalent to the linear term in the antifield expansion of $\Xi$ not contributing to the 
RHS of Eq. (\ref{VariationOfPsiGeneral}).

\refstepcounter{Theorems}
\paragraph     {Theorem \arabic{Theorems}:\label{theorem:VariationOfPsiWithoutMasterEquation}} 
Eq. (\ref{VariationOfPsiGeneral}) actually holds even without assuming that $\rho_{1\over 2}$ satisfies the Master Equation. In 
this case we define $Q_{\rm BRST}$ using the expansion of $\rho_{1\over 2}$ in Darboux coordinates: 
$\rho_{1\over 2} = e^{S_{\rm cl} + Q_{\rm BRST}^a\phi^{\star}_a + \ldots}$.

\paragraph     {Proof:} nothing in the proof of Theorem \ref{theorem:VariationOfPsiGeneral} requires the use of Master Equation.

\paragraph     {Definition} Suppose that $\rho_{1\over 2}$ satisfies the Master Equation sufficiently close to 
the boundary, in the sense that $\Delta_{\rm can}\rho_{1\over 2}$ has compact support which does not touch the 
boundary. Then $Q_{\rm BRST}$ defined as in Theorem \ref{theorem:VariationOfPsiWithoutMasterEquation} becomes a symmetry of $S_{\rm BV}|_L$ sufficiently 
close to the boundary. Let us define $\widehat{Q}_{\rm BRST}\Psi_{{\cal O}}$ via the insertion of the BRST current near 
the boundary:
\begin{equation}\label{insert-brst-current}
\widehat{Q}_{\rm BRST}\Psi_{{\cal O}} \stackrel{\rm def}{=} \int_{\rm\scriptstyle {path\atop integral}} [d\phi] \; \left(\oint_{\rm\scriptstyle {near\phantom{n}the\atop boundary}} j_{\rm BRST}\right) {\cal O} \left.\rho_{1\over 2}\right|_L
\end{equation}
If $\rho_{1\over 2}$ satisfies the Master Equation, then $j_{\rm BRST}$ is conserved and:
\begin{equation}
\widehat{Q}_{\rm BRST}\Psi_{\cal O} \;=\; \Psi_{Q_{\rm BRST}{\cal O}}
\end{equation}

\paragraph     {Proof of Eq. (\ref{DOmegaPsi})}    
Now we are ready to prove Eq. (\ref{DOmegaPsi}). Let us first study the usual ``old'' (not equivariant) form $\Omega$ with the boundary. The derivation parallels the case with no boundary:
\begin{align}
& d\int_{gL} E(d\widehat{g}\widehat{g}^{-1})\;\rho_{1\over 2} \; = \;
\nonumber\\  
\;=\; & \int_{gL}\left( d(E(d\widehat{g}\widehat{g}^{-1}))\,\rho_{1\over 2} \;+\; {\cal L}_{d\widehat{g}\widehat{g}^{-1}}\left(E(d\widehat{g}\widehat{g}^{-1})\rho_{1\over 2}\right) \right) \; =
\nonumber\\   
\;=\; & \int_{gL}\left( - {1\over 2} \{{\cal H},{\cal H}\}E'({\cal H})\,\rho_{1\over 2} \;+\; {\cal L}_{d\widehat{g}\widehat{g}^{-1}}\left(E(d\widehat{g}\widehat{g}^{-1})\rho_{1\over 2}\right) \right) \; =
\nonumber\\   
& \mbox{\tt\small here we use Eq. (\ref{InterpretationUsingLieDerivative}) for Lie derivative with $f({\cal H})=\int E({\cal H})d{\cal H}$}
\nonumber\\   
\;=\; & \int_{gL} \; \mbox{\tt\small Lie derivative of } \rho_{1\over 2} \mbox{ \tt\small along the Hamiltonian flux of }\left.\int d{\cal H} E({\cal H})\right|_{{\cal H} = d\hat{g} \hat{g}^{-1}} \;=\;
\nonumber\\   
\;=\; & Q_{\rm BRST}\left[q\mapsto \int_{gL}\left( \left.\int d{\cal H} E({\cal H})\right|_{{\cal H} = d\hat{g} \hat{g}^{-1}}\;\rho_{1\over 2} \right) \right]
\label{d-omega-with-boundary}
\end{align}
We should choose $E({\cal H}) = \exp({\cal H})$; in this case $\int d{\cal H}E({\cal H}) = E({\cal H}) = e^{\cal H}$.

\subsection{Interpretation of $\Omega$ as an intertwiner in the presence of a boundary}
The equivariant $\Omega^{\tt C}$ is a particular case of $\Omega\langle e^a \rangle$ when $a$ is a solution to Eq.~(\ref{EquationForLogForEquivariant}). 

We interpret the path integral in the theory on $\Sigma\backslash D$, with a boundary
state, as the path integral over the whole $\Sigma$ with insertions ${\cal O}_1,{\cal O}_2,\ldots$ (inside $D$)
determining this boundary state. Then our form $\Omega$ is defined by the path integral 
in the theory on the {\em whole} compact Riemann surface $\Sigma$:
\begin{equation}
\Omega_{{\cal O}_1{\cal O}_2\cdots} = \int_{gL} e^{d\widehat{g}\widehat{g}^{-1}} {\cal O}_1{\cal O}_2\cdots e^a \rho_{1\over 2}
\end{equation}
But we only allow the variations of $L$ which do not change the theory inside the disk $D$.
In other words, we restrict to $d\widehat{g}\widehat{g}^{-1}$ of compact support inside
$\Sigma\backslash D$. We also assume that $a$ also has compact support localized inside $\Sigma\backslash D$. 

Then, as we explained in Section \ref{sec:OmegaIsIntertwiner}, $\Omega$ intertwines $d$ with $\Delta$:
\begin{equation}
d\Omega_{{\cal O}_1{\cal O}_2\cdots} = - \int_{gL}  e^{d\widehat{g}\widehat{g}^{-1}} \Delta_{\rho_{1\over 2}}({\cal O}_1{\cal O}_2\cdots e^a)\; \rho_{1\over 2}
\end{equation}
Since the support of ${\cal O}_1{\cal O}_2\cdots$ is in $D$, and the support of $a$ is 
in $\Sigma\backslash D$, we have:
\begin{align}
d\Omega_{{\cal O}_1{\cal O}_2\cdots} \;=\;& 
- \int_{gL}  e^{d\widehat{g}\widehat{g}^{-1}} \Delta_{\rho_{1\over 2}}({\cal O}_1{\cal O}_2\cdots) e^a\; \rho_{1\over 2} \;-\;
\\    
& \;-\;\int_{gL} e^{d\widehat{g}\widehat{g}^{-1}} {\cal O}_1{\cal O}_2\cdots (\Delta_{\rho_{1\over 2}} e^a)\; \rho_{1\over 2}
\end{align}
The first term should be interpreted as a nilpotent operator $Q_{\rm BRST}$ acting on the inserted
state:
\begin{equation}
\Delta_{\rho_{1\over 2}}({\cal O}_1{\cal O}_2\cdots) = Q_{\rm BRST}({\cal O}_1{\cal O}_2\cdots)
\end{equation}
Therefore we have:
\begin{equation}
d\Omega_{{\cal O}_1{\cal O}_2\cdots}\langle e^a\rangle \;=\; 
- \Omega_{Q_{\rm BRST}({\cal O}_1{\cal O}_2\cdots)}\langle e^a\rangle 
- \Omega_{{\cal O}_1{\cal O}_2\cdots}\langle \Delta_{\rho_{1\over 2}} e^a\rangle 
\end{equation}
\commentstarts{\small
On a compact Riemann surface $(\Delta_{\rho_{1\over 2}}e^a)\rho_{1\over 2}$ is the same as $\Delta_{\rm can}(e^a\rho_{1\over 2})$ because $\rho_{1\over 2}$ satisfies
the Master Equation (see Section \ref{sec:CanonicalOperator}). But in the presence of a boundary, these two expressions
are different, because $\Delta_{\rm can}\rho_{1\over 2}$ produces nonzero boundary terms 
\cite{Cattaneo:2012qu,Cattaneo:2015vsa}. On the other hand, $\Delta_{\rho_{1\over 2}}{\cal O}$ is of compact support if $\cal O$ is of compact support.
}
\commentends

\vspace{7pt}
\noindent
We conclude that the presence of a boundary modifies the intertwiner property 
of $\Omega$. This is a particular case of the following general construction. Let 
$\rho\;:\;\widetilde{\bf g}'\to \mbox{End}(V\,)$ be a representation of of the cone Lie superalgebra. Then any complex 
$K, d_K$ defines a new representation of $\widetilde{\bf g}'$:
\begin{align}
\rho_1\;:\; & \widetilde{\bf g}'\to \mbox{End}(V\otimes K\,)
\\
\rho_1(d) \;=\; & \rho(d) + d_K
\\
\left.\rho_1\right|_{\widetilde{\bf g}} \;=\; & \left.\rho\right|_{\widetilde{\bf g}}
\end{align}
In our case $K$ is the Hilbert space of states on the boundary, with $d_K = Q_{\rm BRST}$.

\subsection{Base form}
In the presence of a boundary our construction of the base form does not give a closed form:
\begin{equation} d\Omega_{\Psi}^{\tt base} = \Omega_{Q_{\rm BRST}\Psi}^{\tt base} \end{equation}
At the same time, $\Omega^{\tt base}$ is still horizontal and invariant. Horizontality follows from the construction, 
as we obtain our base form from the equivariant form by horizontal projection. Invariance follows from 
the fact that $d\Psi^{\tt base}$ is also horizontal.

For our constructions presented in this part, it is important that we restrict to only those variations 
of the complex structure which are zero at the boundary. Otherwise, the variation would change 
the Hilbert space of states. In such case it would be nontrivial to define $d\Psi$, as we would 
need a connection on the bundle of Hilbert spaces.

\section{Unintegrated vertex operators}\label{sec:UnintegratedVertexOperators}
Recall that $H$ is the group of diffeomorphisms of the worldsheet and $\bf h$ its Lie algebra.

\subsection{Definition of unintegrated vertex}
Let us try to relax the condition of the $\bf h$-invariance for the vertex, i.e. do not require 
that $\{\Delta\Phi,U\} = 0$. In this case the deformed action will not be $\bf h$-invariant. It does not 
seem to be possible to modify the definition of ${\cal F}$ so that the deformed action $S + \varepsilon U$ be 
invariant\footnote{Notice however that it is invariant up to BRST-exact terms, because $\{\Delta\Phi,U\} = \Delta \{\Phi, U\}$}.
We will deal with this complication in the following way. Let us assume that, instead of being 
an integral over 
the worldsheet, $U$ is actually a sum of insertions of some operators into fixed points on the 
worldsheet.
We can then restrict the group of diffeomorphisms to a subgroup which preserves those points.
Let ${\bf h}_U$ be the subalgebra of $\bf h$ which preserves the points of insertions and ${\cal F}_U = \Delta^{-1}{\bf h}$.
In other words:
\begin{equation}
\forall \Phi\in {\cal F}_U\;:\;\{\Delta \Phi, U\} = 0
\end{equation}
Remember that we have to integrate $\Omega$ over some cycle in the moduli space of Lagrangian 
submanifolds. We postulate that, in case of unintegrated vertex:
\begin{itemize}
\item    the integration cycle should include the variations of the positions of the insertion 
   points
\end{itemize}
We will now outline the procedure of integration over those insertion points.

\subsection{Integration over the location of insertion points}

\subsubsection{Fixing the Lagrangian submanifold}

Therefore we have to study the restriction of $\Omega$ to the subgroup $H\subset G$. We will for now assume 
that elements of $\cal F$ are in involution ({\it i.e.} $q = 0$ in Section \ref{sec:PropertiesOfF}). We will make use of the 
fact that $H$ (as opposed to full $G$) preserves $\rho_{1\over 2}$. This implies that the integration measure 
can be transformed to a fixed Lagrangian submanifold:
\begin{align}
& \int_{hL}\rho_{1\over 2}\exp\left(dh h^{-1}\right) U\;=
\\
= & \int_{L}\rho_{1\over 2}\;\exp\left((dh h^{-1})\circ h\right) \; U\circ h\label{IntegratedOverFixedL}
\end{align}

\subsubsection{Modified de Rham complex of $H$}
\paragraph     {Definition}
We define the “modified de Rham complex” of $H$ as the space of $H$-invariants:
\begin{equation}
\left(\mbox{Fun}(M \times \Pi T H)\right)^{H}
\end{equation}
where the action of $h_0\in H$ is induced by the right shift on $H$ and the action of $H\subset G$ 
on $M$; in particular, any function of the form $h,x\,\mapsto F(hx)$ is $H$-invariant. The 
differential comes from the canonical odd vector field on $\Pi TH$; we will denote it $d_{(h)}$.
The integrand in Eq. (\ref{IntegratedOverFixedL}) belongs to this space:
\begin{equation}
\exp\left((dh h^{-1})\circ h\right) \; U\circ h \;\; \in\;\; \left(\mbox{Fun}(M \times \Pi T H)\right)^{H}
\end{equation}
We will show that this is the same as the Lie algebra cohomology complex of $\bf h$ with 
coefficients in $\mbox{Fun}(M)$. This is a version of the well-known theorem saying that 
Serre-Hochschild complex of the Lie algebra with trivial coefficients is the same as 
right-invariant differential forms on the Lie group. This is a general statement, true for 
a Lie group acting on a manifold.

\paragraph     {Notations and useful identities}
\begin{align}
\mbox{\tt\small define $\Psi$ so that } \Delta\Psi \;=\; & 
dh h^{-1}= (dhh^{-1})^A{\cal H}_A \mbox{ \tt\small (the moment map)}
\label{ExpandDHHInverse}\\  
\mbox{\tt\small and } \widetilde{\Psi} & = \Psi\circ h\label{def-tilde-psi}
\\
\mbox{\tt\small notice that }\Delta\widetilde{\Psi} & = \widetilde{\Delta\Psi} = (dhh^{-1})\circ h\label{delta-psi-tilde}
\end{align}
In this Section the tilde over a letter will denote the composition with $h$:
\begin{align}
\widetilde{f} = f\circ h \mbox{ \tt\small (a function $x\mapsto f(hx)$)}\label{Tilde}
\end{align}
Here are some identities that we will need: 
\begin{align}
d_{(h)} \Delta\Psi = & -{1\over 2}\{\Delta\Psi,\Delta\Psi\}\label{DHDeltaPsi}
\\
d_{(h)} \Psi = & {1\over 2}\{\Psi,\Delta\Psi\}\label{DHPsi}
\\
d_{(h)} \widetilde{\Psi} = & -{1\over 2}\{\widetilde{\Psi},\Delta\widetilde{\Psi}\}\label{DHtildePsi}
\\
d_{(h)} \Delta\widetilde{\Psi} = & {1\over 2}\{\Delta\widetilde{\Psi},\Delta\widetilde{\Psi}\}\label{DHDeltatildePsi}
\\
d_{(h)} \widetilde{f} = & \{\Delta\widetilde{\Psi}\,,\,\widetilde{f}\} = \{\widetilde{\Delta\Psi}\,,\,\widetilde{f}\} = \widetilde{\{\Delta\Psi\,,\,f\}}\label{d-h-on-f-tilde}
\end{align}
Elements of the space $\left(\mbox{Fun}(M \times \Pi TH)\right)^{H}$ can be obtained from letters $\widetilde{\Psi}$ and $\widetilde{U}$ by operations of multiplication and computing the odd Poisson bracket, or applying $\Delta$.

\subsubsection{Intertwiner between $-d_{(h)} + \Delta$ and $\Delta$}
Consider any function $f\in \mbox{Fun}\left(M\times \Pi TH\right)$ (not necesserily $H$-invariant). We have:
\begin{equation}
(\Delta - d_{(h)})\left(e^{\Delta \Psi\circ h} f\circ h\right) = e^{\Delta \Psi\circ h} \left((\Delta - d_{(h)})f\right)\circ h\end{equation}
(Here $e^{\Delta \Psi\circ h} f\circ h$ is just the product of two functions, $e^{\Delta \Psi\circ h}$ and $f\circ h$.)
In particular, when $f(x,h)$ only depends on $x$ and does not depend neither on $h$ nor on $dh$. 
(i.e. when $f\in \mbox{Fun}(M)$):
\begin{equation}\label{DeltaPsiIsIntertwiner}
(\Delta - d_{(h)})e^{\Delta \widetilde{\Psi}}\widetilde{f} = e^{\Delta \widetilde{\Psi}} \Delta \tilde{f}
\end{equation}
In other words, the operator of multiplication by $e^{\Delta\widetilde{\Psi}}$ intertwines between $\Delta$ and $\Delta - d_{(h)}$. 
After we integrate over the Lagrangian submanifold, $\Delta - d_{(h)}$ becomes just $-d_{(h)}$.

\subsubsection{Integration}\label{sec:IntegrationProcedure}
\paragraph     {The one-form component}
\begin{equation}
\int_L\rho_{1\over 2}\;\Delta\widetilde{\Psi}\; \widetilde{U}\;=\; \int_L\rho_{1\over 2}\;\{\widetilde{\Psi}\,,\,\widetilde{U}\}\;=\;\int_L\rho_{1\over 2}\;\{\Psi,U\}\circ h
\end{equation}

\paragraph     {The two-form component}

\begin{align}
& \int_L\rho_{1\over 2}\;\Delta\widetilde{\Psi}\;\Delta\widetilde{\Psi}\; \widetilde{U}\;=\;
\nonumber\\    
= & \int_L\rho_{1\over 2}\;\{\widetilde{\Psi}\,,\,\{\widetilde{\Psi}\,,\,\widetilde{U}\}\} \;\; - \;\; 2 d_{(h)}\int_L\rho_{1\over 2} \widetilde{\Psi}\widetilde{U}
\end{align}

\subsection{Cohomology of $\Delta$ vs Lie algebra cohomology}
In this Section we will show that the Modified de Rham complex of $H$ is the same as the 
Lie algebra cohomology complex of $\bf h$ with coefficients in $\mbox{Fun}(M)$.

\subsubsection{Definition of the Lie algebra cohomology $H^{\bullet}({\bf h}, \mbox{Fun}(M))$}
The space $\mbox{Fun}(M)$ is a representation of $\bf h$; the action of $\bf h$ is slightly easier to write down at the level of the corresponding action of the Lie group $H$; $h\in H$ acts on $f\in\mbox{Fun}(M)$ as follows:
\begin{equation}
(h.f)(x) = f(h^{-1}x)\,,\;\mbox{\tt\small in other words: }\; h.f = f\circ h^{-1}
\end{equation}
Therefore:
\begin{equation}\label{DLieOnFunctions}
d_{\rm Lie} f = - \{d h h^{-1}\,,\,f\} = - \{\Delta\Psi\,,\, f\}
\end{equation}
We define $d_{\rm Lie}\Psi$ to coincide with $-d_{(h)}\Psi$ of Eq. (\ref{DHDeltaPsi}):
\begin{align}
d_{\rm Lie} \Delta\Psi = & {1\over 2}\{\Delta\Psi,\Delta\Psi\}
\\
d_{\rm Lie} \Psi = & - {1\over 2}\{\Psi,\Delta\Psi\}\label{dPsiAgain}
\end{align}
To follow the Faddeev-Popov notations, we introduce the Faddeev-Popov ghost:
\begin{align}
{\bf c}^A = & (dh h^{-1})^A\label{DLiePsi}
\\  
\mbox{\tt\small then }\Psi = & (-)^{\bar{A}+1}{\bf c}^A (\Delta^{-1} {\cal H}_A) 
\mbox{ \tt\small (cp. Eq. (\ref{ExpandDHHInverse}))}\label{DLieDeltaPsi}
\\  
& {\cal H}_A\in \mbox{Fun(M)}
\end{align}
Beware that $\Psi$ is not just the Faddeev-Popov ghost; it is the product of the Faddeev-Popov ghost ${\bf c}^A$ with $\Delta^{-1}{\cal H}\in \mbox{Fun}(M)$.

Eqs. (\ref{DLiePsi}) and (\ref{DLieDeltaPsi}) are equivalent to saying:
\begin{equation}
d_{\rm Lie} {\bf c}^C = {1\over 2} (-)^{\bar{A}(\bar{B}+1)}{\bf c}^A {\bf c}^B\;f_{AB}{}^C
\end{equation}
where $f_{AB}{}^C$ is the structure constants of $\bf h$:
\begin{align}
\{{\cal H}_A\,,\,{\cal H}_B\} = & f_{AB}{}^C{\cal H}_C
\\
f_{AB}{}^C = & (-)^{\bar{A}\bar{B} + 1} f_{BA}{}^C
\end{align}
It is straightforward to verify using Eqs. (\ref{DLieOnFunctions}) and (\ref{dPsiAgain}) that:
\begin{equation}
d_{\rm Lie}^2 f = 0
\end{equation}
The subgroup $H\subset G$ preserves $\rho_{1\over 2}$ and therefore $\Delta$:
\begin{equation}
\Delta (h.f) = h.\Delta f
\end{equation}
This implies:
\begin{equation}
\Delta d_{\rm Lie} + d_{\rm Lie}\Delta = 0
\end{equation}
\subsubsection{Proof that $d_{(h)}$ is the same as $d_{\rm Lie}$}
This is similar to the statement that for any Lie group $G$, the de Rham subcomplex of right-invariant forms on $G$ is the same as the Lie cohomology complex of $\bf g$ with coefficients in the trivial representation:
\begin{equation}
\left(\mbox{Fun}(\Pi TG) \right)^G = C^{\bullet}({\bf g},{\bf C})
\end{equation}
Our case is a variation on this theme:
\begin{equation}
\left(\mbox{Fun}(M\times \Pi TH)\right)^H = C^{\bullet}({\bf h},\mbox{Fun}(M))
\end{equation}
As we explained, $\left(\mbox{Fun}(M\times \Pi TH)\right)^H$ consists of functions of $\widetilde{\Psi}$ and $\widetilde{U}$. To obtain the corresponding element of $C^{\bullet}({\bf h},\mbox{Fun}(M))$, we replace $\widetilde{\Psi}$ with $\Psi$ with $(dgg^{-1})^A\mapsto {\bf c}^A$ (as in Eq. (\ref{DLieDeltaPsi})), and $\widetilde{U}$ with $U$. Under this identification $d_{(h)}$ becomes $-d_{\rm Lie}$.

\subsection{Another intertwiner between $d_{\rm Lie} + \Delta$ and $\Delta$}
One intertwiner between $d_{\rm Lie} + \Delta$ and $\Delta$ is already provided by Eq. (\ref{DeltaPsiIsIntertwiner}), but it is nonlocal
(because each $\Delta \Psi$ contains one integration). Motivated by the integration procedure of
Section \ref{sec:IntegrationProcedure}, we will now construct another intertwiner, a local one.

Let us assume that elements of subspace $\cal F$ are all in involution, i.e. $q(x,y)=0$. In this case:
\begin{equation}\label{PsiPsiDeltaPsi}
\{\Psi, \{\Psi,\Delta\Psi\}\} = 0
\end{equation}
We denote $e^{\{\Psi,\_\}}$ the following operation:
\begin{align}
e^{\{\Psi,\_\}}\;:\;\;\mbox{Fun}(M) \quad\longrightarrow\quad & \mbox{\tt\small(polynomials of $\bf c$)}\otimes \mbox{Fun}(M)\label{TypeOfTheIntertwiner}
\\
U \quad\mapsto\quad & U + \{\Psi,U\} + {1\over 2}\{\Psi,\{\Psi,U\}\} + {1\over 6} \{\Psi,\{\Psi,\{\Psi,U\}\}\} + \ldots
\end{align}
This operation has the following property:
\begin{equation}\label{ConjugationWithPsi}
(d_{\rm Lie} + \Delta) e^{\{\Psi,\_\}} = e^{\{\Psi,\_\}} \Delta
\end{equation}
The action of $\Delta$ on the left hand side is only on $\mbox{Fun}(M)$ (it does not touch the $\bf c$-ghosts)

\subsection{Descent Procedure}
Here we will show that the intertwining operator of Eq. (\ref{ConjugationWithPsi}) can be interpreted as the 
generalization of the string theory descent procedure, which relates unintegrated and 
integrated vertex operators.

Consider an unintegrated vertex opearator $U$. We interpret it as an element of the cohomology 
of $\Delta$ with some ghost number $n$:
\begin{equation}
U\in H^n(\Delta)
\end{equation}
usually $n=2$ in closed string theory and $n=1$ in open string theory.

Eq. (\ref{ConjugationWithPsi}) allows us to construct from $U$ a cohomology class of $\Delta + d_{\rm Lie}$, where $d_{\rm Lie}$ is the 
Lie algebra cohomology differential of our Lie algebra $\bf h$ with coefficients in $\mbox{Fun}(M)$, 
as follows:
\begin{equation}
e^{\{\Psi,\_\}} U\;\in\;H(\Delta + d_{\rm Lie})
\end{equation}
This expression is “inhomogeneous”, in the sense that different components have different ghost numbers. Each application of $\{\Psi,\_\}$ decreases the ghost number by one, but at the same time rises the degree of the Lie algebra cochain. In the context of closed string, the top component coincides with $U$, then goes $\{\Psi,U\}$, then $\{\Psi,\{\Psi,U\}\}$ and so on. In particular, $\{\Psi,\{\Psi,U\}\}$ is our interpretation of the {\em integrated} vertex operator.

We could have used $e^{\Delta \Psi}$ instead of $e^{\{\Psi,\_\}}$. We prefer to use $e^{\{\Psi,\_\}}$ because it leads to the 
local result. Although $\Psi$ contains integration, the odd Poisson bracket is local 
(i.e. involves a delta-function) and therefore removes the integral.

In string theory the use of such an inhomogeneous expression is often referred to as 
``the descent procedure''.

\subsubsection{Integrated vertex and Lie algebra cohomology}
We have shown that the cohomology of $\Delta$ is the same as the cohomology of $\Delta + d_{\rm Lie}$. The cohomology of $\Delta + d_{\rm Lie}$ can be computed using the spectral sequence, corresponding to the filtration by the ghost number. Let $F^p \subset \mbox{Fun}(M)$ consist of the functions with the ghost number $\geq p$. At the first page, we have:
\begin{equation}
E_1^{p,q} = {\mbox{ker } d_{\rm Lie}\;:\; F^pC^q \to F^p C^{q+1}\over \mbox{im } d_{\rm Lie}\;:\; F^p C^{q-1} \to F^p C^q}
\end{equation}
Therefore, if $E_1=E_{\infty}$, then the cohomology of $\Delta$ is equivalent to the cohomology of $\bf h$ 
with values in $\mbox{Fun}(M)$.

\subsubsection{Comparison of $e^{\{\Psi,\_\}}$ and $e^{\Delta\Psi}$}
We have two operators satisfying the identical intertwining relations:
\begin{align}
d_{\rm Lie} e^{\Delta\Psi} + [\Delta, e^{\Delta\Psi}] = 0
\\
d_{\rm Lie} e^{\{\Psi,\_\}} + [\Delta, e^{\{\Psi,\_\}}] = 0
\end{align}
This suggests the existence of some operator $A$ such that:
\begin{equation}
e^{\Delta\Psi} = e^{\{\Psi,\_\}} + d_{\rm Lie} A + [\Delta, A]
\end{equation}
This $A$ is an inhomogeneous operator-form:
\begin{equation}
AU = \Psi U \;+\; {1\over 2} \Psi\Delta\Psi U + {1\over 2}\Psi\{\Psi,U\} \;+\; \ldots
\end{equation}

\subsubsection{Relation between integrated and unintegrated vertices}
Consider the special case of flat worldsheet. There is a subalgebra ${\bf R}^2\subset {\bf h}$ consisting 
of translations ($\partial\over\partial z$ and $\partial\over\partial \overline{z}$). Let us restrict $h$ to this subalgebra. This simplifies the 
computation because: $H^{n>2}({\bf R}^2,\mbox{Fun}(M))=0$. Therefore we have:
\begin{equation}
e^{\{\Psi,\_\}} U = U + \{\Psi,U\} + {1\over 2} \{\Psi,\{\Psi,U\}\}
\end{equation}
Going back from the Faddeev-Popov notations to the form notations: ${\bf c}^A\mapsto (dh h^{-1})^A$ we 
obtain:
\begin{equation}\label{UsualIntegrated}
\{\Psi,\{\Psi,U\}\} = dz\wedge d\overline{z}\; b_{-1}\overline{b}_{-1} U
\end{equation}
(here $dz$ and $d\overline{z}$ is what remains of $dh h^{-1}$). This is the usual integrated vertex operator 
of the bosonic string theory.

\appendix

\section{Supermanifolds}
\subsection{Contraction and Lie derivative}

We define $\iota_V$ for a vector field $V$ as follows. 
If $V$ is even, we pick a Grassmann odd parameter $\epsilon$ and define:
\begin{equation} 
(\iota_V\, \omega)(Z,dZ) = {\partial\over\partial\epsilon} \omega(Z, dZ +\epsilon V\,) 
\end{equation}
Remember that $dZ$ parametrizes a point in the fiber of $\Pi TM$ over the point $Z$ in $M$. 
Then $dZ + \epsilon V$ is a new point in the same fiber, linearly depending on $\epsilon$.

If $V$ is odd, we define $\iota_V$ as follows: 
$\iota_V\,\omega = {\partial\over\partial \epsilon}\iota_{\epsilon V}\,\omega$. In coordinates:
\begin{equation} \iota_V = V^A{\partial\over\partial dZ^A} \end{equation}
The relation to Lie derivative:
\begin{equation} [\iota_V,d] = {\cal L}_V \end{equation}

\subsection{Symplectic structure and Poisson structure}\label{sec:PoissonStructure}
Consider a supermanifold $M$, with local coordinates $Z^A$, equipped with an odd Poisson 
bracket of the form:
\begin{equation} 
\{F,G\}\;=\; F\stackrel{\leftarrow}{\partial\over\partial Z^A} ^A\pi^B {\partial\over\partial Z^B} G 
\end{equation}
The Poisson bivector $^A\pi^B$ should be symmetric in the following sense:
\begin{equation} ^A\pi^B = (-)^{\bar{A}\bar{B} + \bar{A} + \bar{B}} \;^B\pi^A \end{equation}
The odd symplectic form $\omega$ can be defined from the following equation:
\begin{equation} dF = (-)^{\bar{F} + 1}\iota_{\{F,\_\}} \omega \end{equation}

\subsection{Darboux coordinates}
In Darboux coordinates:
\begin{align}
\{F,G\}\;=\; & F\left( \stackrel{\leftarrow}{\partial\over\partial \phi^{\star}_A}\;{\partial\over\partial\phi^A} \;-\; \stackrel{\leftarrow}{\partial\over\partial \phi^A}\;{\partial\over\partial\phi^{\star}_A} \right)G \label{DarbouxCoordinates}
\\
\omega\;=\; & (-1)^A d\phi^A d\phi^{\star}_A
\end{align}

\section{Proof of the theorem-definition \ref{theorem:DeltaCanonical}}\label{sec:ProofOfTheoremDef}
\paragraph     {Lemma 1} Our $\mu_L[\rho_{1\over 2}]$ (which is a density on $L$ defined, given $\rho_{1\over 2}$, by Eq. (\ref{MeasureMu})) only depends on $\rho_{1\over 2}$ through restriction to the first infinitesimal neighborhood of $L$. In other words, if we replace $\rho_{1\over 2}$ with $e^f\rho_{1\over 2}$ where $f$ is a function on $M$ having second order zero on $L\subset M$, then $\mu_L$ will not change.

\commentstarts{\small
This is slightly counterintuitive, because $\Delta_{\rm can}$ is actually a second order differential operator. It is important that $L$ is Lagrangian.
}\commentends

\noindent
\paragraph     {Proof} The definition of $\mu$ is given by Eq. (\ref{MeasureMu}); $\rho_{1\over 2}$ only enters the left hand side of Eq. (\ref{MeasureMu}) through the first infinitesimal neighborhood of $L$.

\vspace{10pt}
\noindent
We will now prove that a function $\Psi\in C^{\infty}(L)$ can {\em locally} be extended from a Lagrangian
submanifold $L$ into the BV phase space $M$ so that the Hamiltonian vector field of the
extended $\Psi$ preserves $\rho_{1\over 2}$. (This is only true locally.)

\paragraph     {Lemma 2} For any point $x\in L$, a fixed positive integer $n$, and a smooth
function $\Psi$ on $L$, exists an open neighborhood $U\subset M$ of $x$, such that $\Psi$ can be extended
from $U\cap L$ to a function $\widetilde{\Psi}$ on $U$ such that the derivative of $\rho_{1\over 2}$ along the flux of $\{\widetilde{\Psi},\_\}$
has zero of the order $n$ on $U\cap L$.

\paragraph     {Proof} Direct computation in coordinates. Let us choose some Darboux coordinates 
$\phi^i,\phi_i^{\star}$, so that $L$ is at $\phi^{\star}=0$. Let us use these coordinates to identify half-densities with 
functions. Without loss of generality, we can assume that in the vicinity of $m$:
\begin{align}
& \rho_{1\over 2} = e^{S}
\\   
& \mbox{\tt\small where $S=s(\phi) + Q^i\phi^{\star}_i + \ldots$}
\end{align}
where $\ldots$ stand for terms of the higher order in $\phi^{\star}$. Then our problem is to find $\widetilde{\Psi}(\phi,\phi^{\star})$ 
solving:
\begin{align}
& (-1)^{\bar{i}}{\partial\over\partial\phi^i}{\partial\over\partial\phi^{\star}_i}\widetilde{\Psi} + \{S, \widetilde{\Psi}\} = 0
\\   
& \widetilde{\Psi}(\phi,0) = \Psi(\phi)
\end{align}
Solutions can always be found, order by order in $\phi^{\star}$, to any order $n$.
For example, when $n=1$:
\begin{align}
\widetilde{\Psi}(\phi,\phi^{\star}) = \Psi(\phi) + \chi^i(\phi)\phi^{\star}_i
\end{align}
where $\chi^i(\phi)$ should satisfy:
\begin{equation}
\partial_i((-1)^{\bar{i}}e^s\chi^i) = - e^s Q\Psi
\end{equation}
This equation always has a solution in a  sufficiently small neighborhood $U$ of $x$.

\paragraph     {Lemma 3}
\begin{equation}\label{EquivarianceOfMu} g^* \mu_{gL}[\rho_{1\over 2}] = \mu_L[g^* \rho_{1\over 2}] \end{equation}
Proof For any $H\in {\rm Fun}(M)$:
\begin{align}
\int_L (H\circ g) \; g^*(\mu_{gL}[\rho_{1\over 2}])\;=\; &
\\
=\;\int_{gL} H \mu_{gL}[\rho_{1\over 2}] \;=\; & \left.{d\over dt}\right|_{t=0}\int_{e^{t\{H,\_\}}gL}\left.\rho_{1\over 2}\right|_L\;=\;
\\
\;=\; & \left.{d\over dt}\right|_{t=0}\int_L g^*\left(e^{t\{H,\_\}}\right)^*\rho_{1\over 2}\;=\; \left.{d\over dt}\right|_{t=0}\int_L \left(\exp(t\{H\circ g,\_\})\right)^*g^* \rho_{1\over 2}\;=\;
\\
=\;\int_L (H\circ g) \mu_L[g^*\rho_{1\over 2}] &
\end{align}
{\bf Proof of Theorem} We can in any case define $\sigma_{1\over 2}[L,\rho_{1\over 2}]$ by the formula:
\begin{equation}\label{def-sigma-L} \left.\sigma_{1\over 2}[L,\rho_{1\over 2}] \right|_L \;=\; \mu_L[\rho_{1\over 2}] \end{equation}
What we have to prove is that:
\begin{align}\label{sigma-is-L-independent-in-words}
\mbox{\tt\small so defined $\sigma_{1\over 2}[L,\rho_{1\over 2}]$ does not depend on $L$}
\end{align}
Consider any $x\in M$ and a Lagrangian submanifold $L\subset M$ such that $x\in L$ and $e_1,\ldots,f^1,\ldots$ in $T_xM$ such that $e_1,\ldots $ are tangent to $L$ and $\omega(f_i,e^j) = \delta_i^j$. 
Then, Eq. (\ref{def-sigma-L}) says:
\begin{align}
\mbox{\tt\small by definition }\sigma_{1\over 2}[L,\rho_{1\over 2}] (x,\,e_1,\ldots,f^1,\ldots) \;=\; \mu_L[\rho_{1\over 2}](x)(e_1,\ldots)
\end{align}
Let us consider Eq. (\ref{EquivarianceOfMu}) in the special case when $g\in G$ is such that $g(x)=x$. We get:
\begin{equation}\label{SigmaRotated} \sigma_{1\over 2}[gL,\rho_{1\over 2}](x,\, g_*e_1,\ldots, g_*f^1,\ldots)\;=\; \sigma_{1\over 2}[L,g^*\rho_{1\over 2}](x,\,e_1,\ldots, f^1,\ldots) \end{equation}
Consider an infinitesimal variation of $L$ specified by some “gauge fermion” $\Psi\in\mbox{Fun}(L)$. Let us use Lemma 3 to extend it to $\widetilde{\Psi}$, and put $g = \exp\left(t\{\widetilde{\Psi},\_\}\right)$. Lemma 2 implies that $\left.{d\over dt}\right|_{t=0}$ of the RHS of Eq. (\ref{SigmaRotated}) vanishes. This proves that the variation with respect to $L$ of the LHS of Eq. (\ref{SigmaRotated}) vanishes, and therefore $\sigma_{1\over 2}[L,\rho_{1\over 2}]$ does not depend on $L$.

\section*{Acknowledgments}
We are greatful to   Nathan~Berkovits, Alexei~Kotov, Michael~Movshev, Alexei~Rosly and Albert~Schwarz for userful discussions.
This work was partially supported by the FAPESP grant 2014/18634-9 
``Dualidade Gravitac$\!\!,\tilde{\rm a}$o/Teoria de Gauge'', 
and in part by the RFBR grant 15-01-99504 ``String theory and integrable systems''. 
I would like to thank ICTP-SAIFR for their support through FAPESP grant 2011/11973-4.


\def\cprime{$'$} \def\cprime{$'$}
\providecommand{\href}[2]{#2}\begingroup\raggedright\endgroup

\end{document}